\documentstyle[aps,prl,multicol,graphicx,psfig]{revtex}

\begin{document} 
\draft 
 
\author{ 
Giuseppe Foffi, Gavin D. McCullagh, Aonghus Lawlor, Emanuela Zaccarelli 
and Kenneth A. Dawson 
} 
\address{ 
Irish Centre for Colloid Science and Biomaterials, Department of 
Chemistry,\\ University College Dublin, Belfield, Dublin 4, Ireland } 
\author{ 
Francesco Sciortino and Piero Tartaglia} 
\address{ 
Dipartimento di Fisica, Istituto Nazionale per la Fisica della 
Materia,\\ 
and I.N.F.M. Center for Statistical Mechanics and Complexity,\\ 
Universit\`{a} di Roma La Sapienza, 
P.le A. Moro 2, I-00185 Roma, Italy} 
\author{Davide Pini} 
\address{Istituto Nazionale per la Fisica della 
Materia \\ and Dipartimento di Fisica, Universit\`{a} di 
Milano, Via Celoria 16, 20133 Milano, Italy} 
\author{ and George Stell } 
\address{Department of Chemistry, State University of  
New York at Stony Brook, NY 11794-3400, U.S.A.}

\title{ 
Phase equilibria and glass transition in colloidal systems with 
short-ranged \\ attractive interactions. Application to protein 
crystallization.  } 
 
\date{\today} 
 
\maketitle

\begin{abstract} 
We have studied a model of a complex fluid consisting of particles
interacting through a hard core and a short range attractive potential
of both Yukawa and square-well form.  Using a hybrid method,
including a self-consistent and quite accurate approximation
for the liquid integral equation in the case of the Yukawa fluid,
perturbation theory to evaluate the crystal free energies, and 
mode-coupling theory of the glass transition, we determine both
the equilibrium phase diagram of the system and the lines of
equilibrium between the supercooled fluid and the glass phases.  
For these potentials, we study the phase diagrams  for different values of
the  potential range,  
the ratio of the range of the interaction to the diameter
of the repulsive core being the main control parameter.
Our arguments are relevant to a variety
of systems, from dense colloidal systems with depletion forces,
through particle gels, nano-particle aggregation, and globular protein
crystallization.

\end{abstract} 
 
 
 
\begin{multicols}{2} 
\smallskip 
\section{Introduction} 
Recently novel results in the statistical mechanics of the fluid state 
have emerged from a series of studies of the phase diagram of 
particles with a hard core and short-ranged attractive potential. 
At first sight this type of interaction potential would have been 
expected to yield the well known phase diagrams of 
simple solids, liquids and gases. 
Here we shall see that a new scenario emerges.\\
In many practical situations attractive forces that are 
short-ranged compared to the size of the particles arise because the size of
the particles itself is large, whilst the physical forces retain their
typical microscopic range. Thus, in order to model large molecules, 
such as proteins, colloids and nano-particles, 
one often works in a regime where the ratio of the range of
attraction to the size of the repulsive core is small.
This crucial issue is emerging in
the literature in such areas as protein crystallization
\cite{piazza00,mushol95}, dense colloids
\cite{lekkerkerker,verduin95,poon,bartsch}, nano-particle assemblies,
pre-ceramic particle gelation, latex formation, ``buckyballs''
\cite{buckyballs} and many others. In some arenas its relevance
begins to be recognized, while in others it still remains to be
understood. \\ Atoms or molecules interacting via a potential with a
repulsive core and an attractive tail may form gases, liquids and
solids as a function of the temperature and density. The attraction is
responsible for the liquid-gas transition, while the solid is
dominated by the repulsion.
In the low-to-medium range of densities, particle entropy and energy
enter the free energy and compete to produce liquid (energy-favoured)
and gas (entropy-favoured) phases.  Thus, beneath the critical point,
the total system free energy is optimized if the system splits into
two sub-systems, and thereby a sum of two independent
free-energies. One of these free energy terms is dominated by the
energy (liquid) and the other (gas) by the entropy. A homogeneous, unseparated
system, maintained at the average density of the two sub-systems,
would not be able to fully optimize either energy or entropy. This, in
essence, is the cause of typical first-order phase transitions between
a liquid and a gas.
In the liquid, as the density increases the remaining diffusive
motions of the particles, are reduced and their attendant
(configurational) entropy diminishes. Now, the random structure of a
liquid is favored provided overall diffusive particle motion is
possible. When, at higher densities, diffusion is greatly limited, 
the free volume is better utilized by making a regular crystalline
array, and the entropy of the system is increased by the vibrations
of particles within the regular array of a crystal. This is the reason
why hard-sphere fluids begin to form a regular crystal at about $49\%$
volume fraction, and above the freezing transition at $55\%$ have a higher entropy than a
liquid-like structure at the same volume fraction. Indeed, for hard
spheres, at volume fraction between $49\%$ and $55\%$ the system again
phase separates to achieve the optimal value of entropy: a lower
density fluid of $49\%$ volume fraction and a well-packed crystal at
$55\%$.
Nevertheless, it is well known that in some experimental studies 
the system may be unable to 
access the crystalline state rapidly enough, and the disordered super-cooled liquid 
structure {\it freezes} into a glass. For hard sphere fluids, 
this occurs at a volume fraction higher than about $58\%$. 
The formation of the glass is a signal that equilibrium quantities
such as the free energy may not be sufficient, to describe the behavior 
of the system, and that long-lived dynamically
arrested  states may be important under such conditions.
 
It is possible to relate these qualitative comments to recent 
understanding acquired from experiments, simulations and theory. 
Thus, there is a separation of time scales in dense systems, 
including super-cooled liquids, in which particles spend a long 
time trapped by a surrounding cage and thereafter they escape from 
it. In fact, dynamically slowed systems begin to exhibit a plateau 
in the self-correlation function, reflecting the time spent in 
cages, and then a decay ($\alpha$-relaxation) reflecting escape 
from the cage and free configurational motion. 
This approach to the glass transition was formalized 
\cite{debenedetti01} with the idea that motions in dense fluids 
can be divided into intra- and inter-basin motions, where the basins
refer to the multidimensional potential energy as a function of the
particles coordinates. This separation of time scales, and thereby
type of motion, is reasonable for dense systems, and may be used to
justify a conceptual partitioning of the entropy into two parts, a
configurational and a local contribution \cite{ISentropy}. These ideas
can be re-expressed by considering the system, at a fixed average
density, composed of central particles trying to escape their cages
of neighbours, which are themselves fluctuating, and exchanging with
their neighbours. Evidently, the $\alpha$-relaxation process
corresponds, in systems very close to arrest, to the escape of the
central particles from their cages after some time. In the absence of
attractive forces, or for relatively long-ranged ones, the motions of
the cage are restricted by the packing forces.  This picture is
relatively clear, at least in a phenomenological manner, for hard core
spherical particles. For systems that possess a strong repulsion, and
a long-ranged attraction, there are no new special features. Thus both
``central'' and ``cage'' particles remain in their mutual range of
attraction while the structural rearrangement takes place. The
resulting cage breaking can be viewed as being almost the same as that
for hard spheres, but with changed zero of energy.  Even if we
acknowledge that there are effects due to attractions, they can still
be considered to be weak perturbations of the picture arising from the
hard core spherical particles.

 
Now let us turn from the scenario just described, typical of 
hard-sphere particles or particles where the attractions are long-ranged 
compared to the core
size, to situations where attractions play a
principal role. In this case, the freedom of the cage
particles is considerably reduced. Indeed, particles must remain
within a certain distance from each other where substantial attractive
energies are still available, or they lose the advantages of being in
the liquid-like structure. This being so, the cage around the central
particle is much more rigid, since only smaller excursions from
average positions of these cage particles are possible. In fact, for
sufficiently narrow wells, the time
that a particle spends inside the cage increases; a plateau regime results,
and the system will eventually freeze. This ``cage rigidity'' was also
the determining factor in hard core particles at very high density,
leading to the typical colloidal ``repulsive glass''. Here, though the
mechanism and detailed laws will be different, we see that
short-ranged attractions are able to cause cage rigidity, and
formation of a solid, either a glass, or a crystal.
Following these arguments 
we see that for particles with short-ranged attractions,
the density and temperature window over which the liquid
state is stable is greatly reduced since the configurational entropy
is reduced.  The system separates into a low-density fluid state where
there is sufficiently free motion, and a dense state where there is in
any case not sufficient configurational entropy to sustain a liquid,
and the system freezes. This state will be a crystal or a dynamically
arrested state. Based on the developing views of others, and what we
shall present here, all these expectations are borne out from precise
calculations, and increasingly in experiments.
 
%
From the discussion above, we may expect that when attractions are
short-ranged the arrested glass is of different nature than the
typical repulsive one, since it is favoured by both the energy (because
the particles are not close enough to sample mainly repulsive energy),
and the local entropy (given that the density is yet relatively low).
We have earlier called this arrested state an attractive glass
\cite{A4,zaccarelli01}, and the crystal, previously found by a number
of others \cite{iso-frenkel,tejero,evans}, we name the attractive
crystal. This  distinguishes it from the typical face centred cubic (FCC) 
crystal formed by repulsive forces and the analogous repulsive glassy state.
 
Given the possibility that for short-ranged attractions, solids, 
both crystals and glasses, can be formed by these two distinct 
mechanisms leading to cage-rigidity, we may suppose that it is possible in 
principle for them to co-exist. For fixed (short) ranged potential 
one way of changing the balance of attraction and repulsion is to 
change the density, and we comment that such co-existence has been 
previously shown for crystals \cite{iso-frenkel,tejero,evans}, and 
glasses \cite{A4,zaccarelli01}, and is reproduced here. 
 
In summation, we expect that, as the well width is narrowed, the
liquid state becomes progressively less favoured, and is replaced by a
conventional repulsive crystal, or its equivalent glass. An attractive
crystal, and its equivalent repulsive glass, should also be present at
higher densities, both types of solid co-existing at some typical
densities where attractions and repulsions compete.
From the preceding considerations we expect that these predictions 
should be general, irrespective of the detailed shape of the 
potential, and reflecting the typical range of the potential. 
 
We have studied two potential energy models typically used to mimic
colloidal interactions, the square well (SW) and the hard-core Yukawa
potential, using a variety of techniques of condensed-matter
theory. We have determined the phase diagrams calculating the liquid
free energy using perturbation theory for the SW and the
self-consistent Ornstein-Zernike approximation (SCOZA) for the Yukawa
potential. The crystalline free energy has been calculated applying
second order perturbation theory for both potentials.  This technique
has been applied to short-ranged potentials before
\cite{evans,gast83&86,hagen94}, with quite remarkable success in
reproducing the crystalline free energy in comparison to Monte Carlo
experiments. However, it has been noted
\cite{evans} that the same method is not so satisfactory in
calculating the free energy of the liquid and gas states for very
narrow attractive ranges, and this affects the accuracy of the phase diagram 
\cite{evans,private-evans}. To avoid this problem
in the case of the Yukawa potential, we have modified the calculation
so that the proven good features of the perturbation theory of the
crystal are combined with the proven quality of the SCOZA for the
liquid and gas free energies to produce a phase diagram that is of
uniformly good quality.  Currently there is no working SCOZA method for the SW
case, though there is no fundamental barrier to developing one.
 
In our work  the ideal Mode Coupling Theory (MCT) \cite{goetze91}
of super-cooled liquids has been used in order to locate the glass
transition curves. For colloids this method has also been found to
describe many elements of the transition to the arrested
state \cite{bergenholtz99,fabbian99}, though it does suffer from some
limitations, a matter to which we return later.  In the  MCT
calculations we have used the structure factors from Percus-Yevick
approximation (PYA) for the case of SW and from SCOZA for the case of the
Yukawa potential.  These  have proved to be quite accurate
theories of the liquid and fluid states. The SCOZA method in
particular has been compared to Monte Carlo simulations for a range of
screening parameters (``well-widths'') of the Yukawa potential, and it
has been shown \cite{pini3,caccamo99} that the 
agreement for the phase diagrams is quantitative, at least for modest values 
of screening parameter.

By combining the results from the different methods described above,
we are able to give, for some regimes, what we believe to be quite
accurate phase diagrams. For the general case we believe that the
results are at least qualitatively correct, and provide us a coherent
picture of the connection between well-width and the arrangement of
gas, liquid, crystal and glass phases. It is this overview of how the
various phenomena fit together that is currently missing, and that
should prove useful in the various applications alluded to above.  We
also point out that, from experiments, it has become clear that the
interaction between globular proteins, in the range where they may
crystallize, is characterized by short range attractions
\cite{piazza00}. In the case of proteins, a better comprehension of
the phenomena would imply the study of anisotropic types of potential
\cite{sear99}, due to hydrophilic-hydrophobic patching of the
protein surface. The usual approach is, however, to use an effective
isotropic interaction obtained by averaging over the anisotropy. Thus,
as discussed later, most of our conclusions will be also important to
understand the process of protein crystallization.  

The paper is organized as follows. In the next section \ref{sec:S} we
describe the approximate closures to the Ornstein-Zernike liquid integral
equation that we used for the Yukawa and SW potentials. The methods 
employed to determine the equilibrium
fluid and solid phases are described in sections \ref{sec:solid} and 
\ref{sec:Phasediayuk}. Section \ref{sec:MCT} is devoted to a brief 
sketch of MCT applied to attractive potentials,
while in section \ref{sec:spindal} the relevance
of the spinodal curve for colloidal systems is discussed.  
The complete phase diagram, including the
structural arrest lines is discussed in section \ref{sec:results},
while section \ref{sec:conclusions} is devoted to our conclusions.
 
\section{Approximations to the OZ equation for the Yukawa and 
Square Well models}
\label{sec:S} 
In this section we shall discuss the theory used in the investigation 
of the phase diagram.  The Ornstein-Zernike (OZ) equation for the pair 
correlation function $h(r)$ is 
\begin{equation} 
         h(r) = c(r) + \rho \int d{{\bf r}^{'}} 
         c(|{\bf r}-{{\bf r}^{'}}|) h(|{{\bf r}^{'}}|) 
\label{OZ} 
\end{equation} 
where $g(r) = h(r) + 1$ is the radial distribution function and $c(r)$
the direct correlation function.  Another important quantity is the
static structure factor $S_q$, which is the equal time correlation
function of the density variables in wave vector space,
\begin{equation} 
\label{Sq} 
S_q=\langle \rho_{-q}(t)\rho_{q}(t)\rangle/N 
\end{equation} 
where the average $\langle \cdots \rangle$ is performed at equilibrium and
 the density variables are $\rho_q(t)=\sum_i e^{i {\bf q}\cdot {\bf
r}_i(t)}$, where the sum runs over all $N$ particles in the system. The
Fourier transform of the  correlation function $h_q$ is related to
this quantity by the relation $S_q=1+\rho h_q$. The OZ relation in the
wave vector space reads
\begin{equation} 
\label{OZ-q} 
S_q=\frac{1}{1-\rho \hat{c}_q} 
\end{equation} 
$\hat{c}_q$ being the Fourier transform of the direct correlation function. 
 
As it stands Eq.~(\ref{OZ}) is not closed and some type of
approximation is needed in order to solve it. In what follows we
have chosen to calculate structural and thermodynamical
properties using the PYA for the SW potential and the SCOZA
\cite{pini3,caccamo99,pini01} for the Yukawa potential. Both these model potentials
possess some fundamental properties that make them good candidates
for studying properties of attractive colloidal particles when the
range of interaction is short.  The use of SCOZA has been justified by
the success of such approach in predicting simulation data.
%
\subsection{The SCOZA for the Yukawa potential} 
\label{scoza-sec} 
The application of SCOZA to a hard-core Yukawa fluid provides a 
semi-analytic calculation of the thermodynamic properties of the 
fluid, liquid and gas states of the system 
\cite{pini3,caccamo99}. SCOZA has been 
applied to Yukawa systems with relatively large values of the 
range of the potential, with a satisfactory reproduction of the 
liquid-vapour binodal curves and a good description of the critical 
point region. 
 
The hard-core Yukawa fluid is described by the following 
inter-particle potential, 
\begin{equation} 
\label{eq:hcy}
v(r) = \left\{ \begin{array}{cc} 
             \infty \ \ \ \ \ \ \ \ \ \ \ \ \ \ \ \ \  r <\sigma  \\ 
     \mbox{}-\sigma \epsilon \, {\displaystyle \frac{{\rm e}^{-b (r-\sigma)}}{r}} \ \ \ \   r \geq \sigma 
            \end{array} \right. 
\end{equation} 


The parameter $\epsilon$ defines the energy scale, while
the parameter $b$, known as screening parameter, determines the range
of the potential. The larger the $b$, the shorter is the range of the
potential. In
this paper we set $\sigma = 1$ and $\epsilon =1$, therefore the
screening parameter is in units of the reciprocal of the hard-core
diameter, and the temperature in units of the well depth. 
%

SCOZA provides a closure relation for the OZ equation (\ref{OZ}) by
expressing the direct correlation function $c(r)$ in terms of the
potential $v(r)$, as for other approximations like the mean spherical
approximation (MSA), PYA, hypernetted chain (HNC) approximation etc.,
but introducing one or more state dependent parameters that can be
adjusted to force the system to satisfy various exact thermodynamic
relations of the system. In particular, the simplest SCOZA studied so far
assumes 
\begin{equation}
g(r)=0 \ \ \ \ \ \ \ \ \ \ \ \ \ \ \ r \le \sigma
\label{eq:core}
\end{equation}  
while $c(r)$ for $r\ge \sigma$ is
composed of two contributions, describing respectively the soft part
of the potential and the hard core, as
\begin{equation} 
       c(r) = - A \beta v(r) + K_{HS} {{\rm e}^{-b_{HS} (r-\sigma)}\over{r}} 
       \ \ \ \ \ \ \ \ r \ge \sigma. 
\label{eq:c} 
\end{equation}

The Yukawa function in Eq.~(\ref{eq:c}) takes into account the
contribution to $c(r)$ arising from the hard-core repulsion. Thus, the
two parameters $K_{HS}$ and $b_{HS}$ can be determined by setting
$v(r) = 0$ in Eq.~(\ref{eq:c}) and requiring that both the
compressibility and the virial route to thermodynamics lead to the
Carnahan-Starling equation of state for a hard-sphere fluid
\cite{carnahan69}. This amounts to describing the hard-sphere correlations
via the Waisman parameterization~\cite{waisman}. The soft part of the contribution in (\ref{eq:c}) is
assumed to be proportional to $v(r)$ and hence it has the same range
of the potential. The proportionality constant $A$ is calculated by
imposing the condition that the compressibility and the energy routes
yield the same result.  This corresponds to the condition \cite{caccamo99},
\begin{equation} 
\label{eq:comp} 
       -\frac{\partial}{\partial \beta} \hat{c}(q=0) = 
        \frac{\partial^2}{\partial \rho^2}\left(\frac{U^{ex}}{V}\right)_T 
\end{equation} 
where $U^{ex}$ is the excess internal energy \cite{hansen86}.  Eq.
\ref{eq:comp} implies a partial differential equation (PDE) for 
$A(\rho,\beta)$. 
The Yukawa potential~(\ref{eq:hcy}) lends itself particularly well to the 
SCOZA scheme, because for this kind of interaction it is possible by means 
of Eqs.~(\ref{eq:core}), (\ref{eq:c}) to
establish an analytic relation between $\hat{c}(q=0)$ and $U^{ex}$, which
allows us to obtain straightforwardly a closed PDE from Eq.(\ref{eq:comp}) 
by using $U^{ex}$ as the unknown quantity instead of $A(\rho,\beta)$ 
\cite{pini3,pini01}. Once the internal
energy has been obtained by numerical solution of this PDE, the
Helmholtz free energy is calculated by integration with respect to $\beta$.

\begin{figure}

\centerline{\psfig{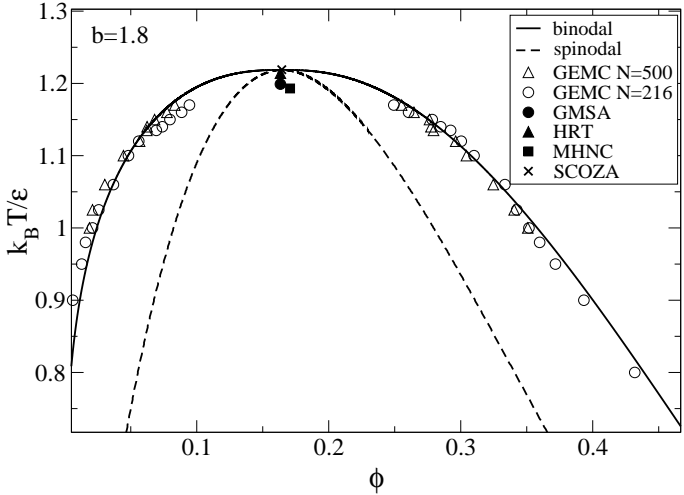}}
\caption{Liquid-liquid phase diagram from SCOZA compared to simulation data for b=1.8. The GEMC data are taken from Shukla \protect\cite{shukla00} (see text for details).  The estimation of the critical points are from generalized means spherical approximation (GMSA), modified hypernetted chain (HNC) and HRT}
\label{fig1.8phdiag}
\end{figure}

\begin{figure}
\centerline{\psfig{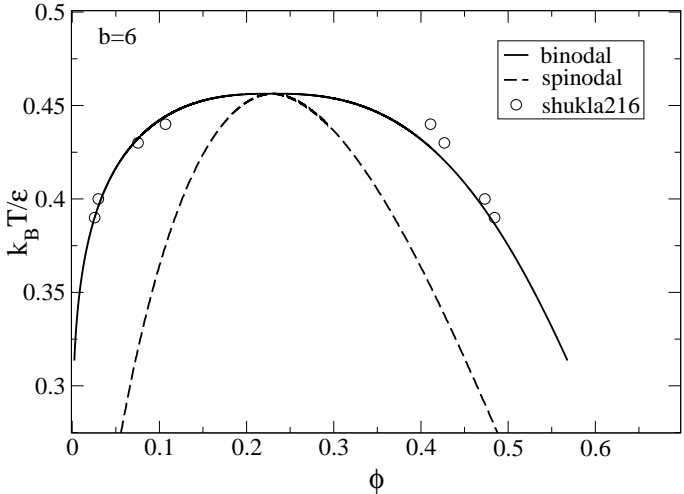}}
\caption{Liquid-liquid phase diagram from SCOZA compared to simulation data for b=6. The GEMC data are taken from Shukla for $N=216$\protect\cite{shukla00} (see text for details).}
\label{fig6.0phdiag}
\end{figure}

The reasons which led us to adopt SCOZA can be summarized as follows.
The method uses a reasonable choice of functional
relationship between $c(r)$ and $v(r)$ on the basis of numerous
calculations obtained over a number of years.  Furthermore, it has
been shown \cite{caccamo99} that the results obtained are in excellent
agreement with computer simulations, at least for not too narrow
ranges of the potential. In this sense it may be viewed as the best
semi-analytical method to study the Yukawa potential.  Finally, it is
simple and convenient to obtain numerical solutions of
Eq.~(\ref{eq:comp}), and this is helpful to make a survey of a problem in
a large parameter window, rather than just in a small part of the
parameter space. To give the reader some sense of the accuracy of
SCOZA in some of the regimes we will discuss later, we present in
Fig. \ref{fig1.8phdiag} and \ref{fig6.0phdiag} the phase diagrams for
$b=1.8$ and $b=6$ calculated using SCOZA, compared with Gibbs Ensemble
Monte Carlo (GEMC) studies \cite{hagen94,shukla00}.  The data by
Shukla are the largest system sizes with this potential yet studied
by GEMC. In all cases, the agreement with SCOZA is excellent, and we
may, in this regime of $b$-values and in calculating the
phase diagrams, consider SCOZA to be equal to the best simulations. 
The method is  superior to the other simple
closures, including for example MSA.

Having said this, we note that there is little real information
available about the detailed reliability of the closure relation where
the potential becomes much narrower than for $b=9$, though it is
reasonable to suppose that many properties are still satisfactory for
somewhat larger $b$-values. Another comment we may make is that, when
the range of the potential narrows, 
Eq.~(\ref{eq:c}) may not be the optimal closure to ensure
that $c(r)$ is accurate. Both these points should be regarded as words
of caution, and as potential directions to develop the SCOZA method to
better represent the structure in such problems. Indeed this will be
the subject of future work. 
 
\subsection{The Percus-Yevick closure for the SW potential} 
\label{sec:sw}
We shall now briefly discuss the closure for a fluid of colloidal 
particles interacting via a square well potential,
\begin{equation}
\label{SWpot}
v(r)=\left \{\begin{array}{cc} \infty \ \ \ \ \ r < \sigma\\ \mbox{}- \epsilon
\ \ \ \ \ \sigma <r < \sigma+\delta
\\ 0  \ \ \ \ \ \  \sigma+\delta < r \end{array} \right. 
\end{equation}
where, in the present discussion, we set $\epsilon=\sigma=1$ and we
define the square well parameter $\Delta=\delta/(\sigma+\delta)$,
which parameterizes the attractive range of the potential. This model
has been already the object of great interest in colloidal science
\cite{rascon,A4}. 
The state of the system is specified by three control parameters, the
packing fraction $\phi= \pi \rho \sigma^3/6$ ( where $\rho$ is the
number density, i.e. $\rho=N/V$), the temperature $k_B T$, and the
square-well parameter $\Delta$ of the attractive shell.  

The PYA for $c(r)$ is $g(r)=0$ for $r< \sigma$ and
\begin{equation} 
c(r) = g(r) \left[1-{\rm e}^{\beta v(r)}\right] 
\end{equation} 
outside the hard core \cite{PY}. We solve the OZ equation in PYA using
Baxter's method of the Wiener-Hopf factorization
\cite{hansen86,baxter68}.  This correspond to rewriting Eq.~(\ref{OZ})
in terms of the real factor function $Q(r)$, defined for $r>0$. For
$0\le r\le R$, $R$ being the range of the potential
($\sigma+\delta$ in the present case), one has
\begin{equation} 
\label{eq:WH1-b}
       r c(r) = - Q'(r) + 2 \pi \rho \int_r^R {d}s\, Q'(s)Q(s-r)\,
\end{equation} 
as well as, for $r>0$, 
\begin{equation}\label{eq:WH2-a}
r h(r) = - Q'(r) + 2 \pi \rho \int_0^R {d}s \, (r-s) h(|r-s|) Q(s)\,. 
\end{equation} 

$Q(r)$ determines $S_q$ via its Fourier transform, 
\label{eq:Sq}
\begin{eqnarray} 
\label{eq:Sq-a}
       S_q^{-1}   &=& \hat{Q}(q)\hat{Q}(q)^*\,,\\\label{eq:Sq-b}
       \hat{Q}(q) &=& 1 - 2 \pi \rho \int_0^\infty {d}r \,  {\rm e}^{i q r} 
Q(r)\,. 
\end{eqnarray} 
%
The resulting equations, obtained implementing the PYA in
Eqs.(\ref{eq:WH1-b}) and (\ref{eq:WH2-a}), are then solved numerically
to calculate the structure factor $S_q$ (see reference \cite{A4} for
more details).
 
\section{The Perturbation Theory Applied to Solid and Fluid phases} 
\label{sec:solid} 
%
\subsection{The Crystal Phase} 
In this section we shall discuss the method used to calculate the
solid free energy.  The perturbative approach has previously been used
by Gast {\it et al.} \cite{gast83&86} to construct the phase diagram
of a colloidal solution with depletion interactions.
 
Further applications of the method appear in references 
\cite{evans,hagen94}. For example, in \cite{evans} the authors
compared the results obtained for both phase diagrams and free
energies of the solid and liquid phases for an Asakura-Oosawa
potential and made comparisons to Monte Carlo calculations. They noted
that the phase diagrams are overall reasonably satisfactory, but that the
crystal free energies are excellent, in most regimes being essentially
quantitative. It was concluded that the potential limitations in
accuracy of the phase diagrams arise from the use of perturbation
theory to the fluid and liquid phases, rather than the crystal. Also,
for the crystal free energies there does not appear to be a
significant loss of accuracy when the range of the potential becomes
narrow.  We have, therefore, applied perturbation theory to the
crystal state of the Yukawa potential.  Besides what we show here, it
is opportune to note that there are in the literature other
perturbative approaches, such as in
\cite{rascon}.
 
The method is summarized as follows. We separate out explicitly the
interaction potential as a hard-core contribution plus the
attractive tail.  The hard-core part of the potential is used as a
reference for the perturbation, and the attractive tail is the
perturbation itself \cite{hansen86}. In other words,  we decompose the
potential as,
\begin{equation} 
\label{pote-div} 
       v(r) = v_{0}(r) + v_{att}(r) 
\end{equation} 
where $v_0(r)$ is the hard-core repulsive potential and expand around
the reference state $v_{0}(r)$.  With this choice, the zeroth order
term of the free-energy expansion coincides with the hard-sphere free
energy.  Once the perturbation expansion is carried out to second
order we have for the Helmholtz free energy the following expression
\cite{hansen86},
\begin{eqnarray}
\label{pertu}
\frac{\beta F}{N} = \frac{\beta F_0}{N} +
\frac{\beta \rho }{2}\int v_{att}(r) g_0(r) d{\bf r} + 
\frac{\beta F_2}{N}
\end{eqnarray}
here  $\beta F_2/N$  is the second-order perturbation term.
$F_0$ and $g_0$ are respectively the Helmholtz free energy and the
radial distribution for the reference hard-sphere system.

We focus our attention on the second order term in the expansion in
Eq. (\ref{pertu}). Indeed, its exact evaluation requires the calculation
of higher order distribution functions \cite{hansen86}, which are very
hard to compute or to approximate reasonably. Barker and Henderson
\cite{barker} proposed an approximation to $F_2$, based on the
following observation.  Since,
\begin{equation}
\label{w}
\frac{\beta F_2}{N}=-\frac{1}{2} \beta (\langle W^2_N \rangle_0 - 
\langle W_N \rangle^2_0)
\end{equation}
where $W_N=\sum_{i<j}^N v_{att}(|{\bf r}_i-{\bf r}_j|)$, Barker and
Henderson proposed to divide the space into concentric spherical
shells and to calculate averaged properties using the number of
particles in each shell. Following this route, they rewrote Eq.\ref{w}
in terms of average numbers in the shells,
\begin{equation}
\label{N}
\frac{\beta F_2}{N}=-\frac{1}{2} \beta \sum_{ij} (\langle N_i N_j\rangle - 
\langle N_i \rangle\langle N_j \rangle) v^i v^j
\end{equation}
where $N_i$ is the number of particles in the shell $i$ and $v_i$ is
the perturbation energy, considered constant, within the shell.  The
first approximation consists of ignoring the correlations between
shells, i.e.
\begin{equation}
\label{app1}
\langle N_i N_j\rangle - 
\langle N_i \rangle\langle N_j \rangle = 0
\end{equation}
for $i \neq j$.
Moreover, inside a given shell, a second approximation is made,
\begin{equation}
\label{app2}
\langle N_i^2\rangle - 
\langle N_i \rangle^2 \approx \langle N_i\rangle k_B T 
(\partial \rho /\partial P)
\end{equation}
The two approximations (\ref{app1}) and (\ref{app2}) are equivalent to
considering the volume of the shells to have the compressibility
properties of a macroscopic portion of space (for more details see
\cite{barker}).  As a result, Barker and Henderson approximated the
second order term in the expansion as,
\begin{equation}
\label{F2}
\frac{\beta F_2}{N}=-\frac{\beta \rho}{4}\left(\frac{\partial \rho}
{\partial P}\right)_0 \int v_{att}^2(r) g_0(r) d{\bf r} 
\end{equation}
This approximation was found to be satisfactory in all calculations
carried out so far. In our work the integrals in Eqs.~(\ref{pertu})
and (\ref{F2}) have been performed by a five-point integration rule,
while for differentiation a central-difference scheme has been used.
\cite{numrecipes}.

To carry out the calculation, we require the Helmholtz free energy and 
radial distribution
function of the unperturbed hard-sphere system in the solid phase. It has been shown by
computer simulation that a hard-sphere fluid shows a solid-fluid
transition, for which the fluid phase alone exists up to a packing
fraction $\phi=0.49$ and the solid FCC phases exists for $\phi>0.55$
\cite{alder}. In between, there is a two phase coexistence of solid and
fluid. These properties are well studied and the various information
required in perturbation theory can be deduced from these studies.  We
note in passing that recently a renewed interest has been arisen in
the equilibrium structure of a hard-sphere crystal. Indeed it has been
believed for a long time that hard spheres crystallize with an FCC
structure. Confocal microscopy observations, however, have rather
found a random hexagonal phase which consists of a stack of FCC and
hexagonal close packing (HCP) layers
\cite{gasser01}. Simulation seems to explain the phenomenon in terms
of the small free-energy difference between FCC and HCP structures
\cite{auer01}. In this paper we assume the crystal equilibrium
structure to be FCC since this is believed to be the more stable \cite{pronk01}. 

To provide continuity with previous authors we make the 
choices described below.  The state equation for a hard sphere FCC solid
has been proposed by Hall \cite{hall} who derived a phenomenological
expression based on computer simulation results, i.e.
\begin{eqnarray} 
\label{Zhall} 
&&Z_{HS} =\frac{P_{HS}V}{Nk_BT}=\\ 
&&\frac{1+\phi+\phi^2-0.67825 \phi^3-\phi^4-0.5\phi^5-6.028 
          \phi^6f(\phi)}{1-3\phi+3\phi^2-1.004305\phi^3} 
\end{eqnarray} 
with $f(\phi) = \exp((\pi \sqrt{2}/6-\phi)[7.9-3.9(\pi 
\sqrt{2}/6-\phi)])$. 
The compressibility can be derived by differentiating the 
compressibility factor $Z_{HS}$ as, 
\begin{eqnarray} 
\label{compre} 
       \left(\frac{\partial \rho}{\partial p} \right)_{0} = 
       \frac{\beta}{Z_{HS}+\phi \left(\partial Z_{HS}/\partial \phi\right)} 
\end{eqnarray} 
In order to calculate the excess hard-sphere Helmholtz free energy 
$F^{ex}_0$ from the compressibility factor $Z_{HS}$, a thermodynamic 
integration in the packing fraction $\phi$ can be performed, obtaining 
\begin{equation} 
\label{the_int} 
\frac{\beta F^{ex}_0}{N}(\phi) =  \frac{\beta F^{ex}_0}{N}(\phi^*)+ 
\int_{ \phi^*}^\phi \left(Z_{HS}-1\right)\frac{d \eta'}{\eta'} 
\end{equation} 
Since the zero-density limit of a FCC crystal cannot be represented as 
easily as the one for a gas, alternative routes to perform the 
thermodynamic integration in (\ref{the_int}) have to be devised 
\cite{alder}. We have chosen to perform the integration starting from 
a packing fraction value of $\phi^*=0.544993$, for which the value for
the free energy has been calculated by computer simulation to be
$F_0^{ex}(\phi^*)\ N = 5.91889$ \cite{frenke-pri}.  We recall that the
excess Helmholtz free energy is defined as the excess with respect to
the ideal gas contribution \cite{hansen86}.

For the radial distribution function $g_0(r)$ we have used an 
analytic formulation proposed by Kincaid and Weis that fits Monte 
Carlo simulation for a hard sphere FCC solid \cite{kinc}. This 
formulation is known to provide a good estimate of the 
hard sphere radial distribution function, at least in the range $0.52 \leq 
\phi \leq 0.56518$. Eqs.~(\ref{pertu}) and (\ref{F2}) can now be solved.

Once the Helmholtz free energy is evaluated, 
following the route we have just described, the 
Gibbs free energy and the pressure can be calculated as

\begin{eqnarray} 
    \beta \! G &=& \frac{\partial \left(\rho \beta \! 
F\right)}{\partial \rho} \label{g}\\ 
    \beta \! P &=& \frac{\rho \beta\! G}{N} -\frac{\rho \beta\! F}{N} 
\label{pre}
\end{eqnarray} 

We have earlier noted that the perturbation theory for the crystal is
highly accurate. Indeed, the second order perturbation term is useful,
but it is interesting to note that the great bulk of the free energy
correction for narrow well problems is captured by the first order
term alone. To understand this, it is worth reflecting on the fact
that the free energy of the crystal in the presence of short-ranged
attractions is referred to the hard core crystal, and there is no
question of the perturbation theory having to determine a priori any
gross structural information. The corrections from attractions arise
by virtue of the small changes in local vibrations that the particles
make around their lattice positions, a portion of these motions
involving the particles being within their mutual attractive range. In
first order perturbation theory these contributions are treated as if
the nature and distribution of the vibrations is unchanged, and the
additional attractive energy contributions calculated essentially as
an integral over the attractive potential multiplied by the
zeroth-order hard core correlation function. The fact that second
order contributions are typically small is suggestive.  In a heuristic
manner, we may argue that the intrinsic limitation on the extent and
complexity of the phase space of the localized particles, imposed by
their being in a crystalline state, means that even when attraction is
incorporated, the changes in the nature and distributions of these
vibrations is small.  We note that this rationale is clearly
inapplicable to the case of liquid, gas, and fluid states, where the
addition of attractions to significantly affect the distribution of
particles motions. Possibly this is the reason why perturbation
theory works well for the crystal, indeed far outside its expected
limitations, but is less successful for the other states. For
completeness we note that for the crystal (Yukawa and SW) the errors, as
estimated by the ratio of second-order to first-order terms are typically
of order $0.5\%$. For the liquid they can be larger. We have however, also studied a square-well fluid using
perturbation theory, and used these results along with those for the
crystal to generate a phase diagram. The perturbation theory of the
square-well fluid is therefore briefly discussed below.
 
\subsection{Liquid phase for the SW fluid} 
In this section we shall discuss the method we adopted to calculate 
the thermodynamical properties of a fluid of colloidal particles 
interacting via a SW potential. 
 
We chose the hard-sphere fluid as the reference system and  
treated the attractive part as the perturbation. The natural 
choice to describe the thermodynamics of a hard-sphere fluid is 
the Carnahan-Starling (CS) equation of state \cite{carnahan69}, 
\begin{equation} 
\label{CS} 
       \frac{\beta P}{\rho} = \frac{1+\eta+\eta^2-\eta^3}{(1-\eta)^3} 
\end{equation} 
The CS equation provides an accurate account of the
thermodynamic behaviour of the hard-sphere fluid for the entire
region of the fluid phase. Its very simple analytical form makes it
possible to obtain a closed expression for the Helmholtz free energy
by integrating over density, as in Eq.~(\ref{the_int}). 
The zero-density limit of the free energy  is the ideal
gas value, so the thermodynamic integration starts from zero
density. Thus, we obtain
\begin{equation} 
\label{Free_CS} 
\frac{\beta F^{ex}}{N} = \frac{\eta (4-3 \eta)}{(1-\eta)^2} 
\end{equation} 
The compressibility is evaluated as in Eq.~(\ref{compre}) by
differentiation.  For the radial distribution function we use a
modification of the analytical PYA $g_0(r)$ for a hard-sphere fluid
that was proposed by Verlet and Weis \cite{VW} to overcome certain
limitations of this closure. Indeed, in PYA \cite{PY}, the contact
value $g_0(r\!=\!\sigma)$ of the radial distribution function
underestimates the real value obtained by computer simulation and,
also, the oscillations of the tail are slightly out of phase and too
weakly damped. Verlet and Weis proposed,
\begin{equation} 
\label{vw1} 
g_0(r/\sigma, \phi) = g'_0(r/\sigma', \phi')+\delta g_1(r) 
\end{equation} 
Here, $g(r)$ is written as the sum of two terms. The first term
corresponds to the solution for $g_0(r)$ within PYA, but evaluated at
a smaller packing fraction value $\phi'$ and a smaller diameter
$\sigma'$, while the second term is a short-ranged correction $\delta
g_1(r)$. The parameter $\sigma'$ is then evaluated via a minimization
of the difference between the simulation result for $g_0(r)$ and the
PYA analytical result, between $1.6\sigma$ and $3\sigma$. This
contribution improves the long-range behaviour of the PYA result. The
addition of the short-range term $\delta g_1(r)$ improves the value at
$r=\sigma$. Analytical forms for $\phi'$ and $\delta g_1(r)$ are given
by the authors \cite{VW}. The improved radial distribution is within
$1\%$ of the computer simulation result in the whole range of packing
fractions.  With this result, the Helmholtz free energy for the 
hard-sphere reference system is calculated, and we may then proceed as in
the previous section to calculate the Gibbs free energy $G$ and
pressure $P$ using Eqs.~(\ref{g}) and (\ref{pre}).

\section{The Construction of Phase Diagram of  
the Yukawa Model by Hybrid Method} 
\label{sec:Phasediayuk} 

In the case of the Yukawa potential we have modified the approach of
previous researchers somewhat in order to obtain the benefits of the
best methods of condensed and liquid state theory. We have used SCOZA
to calculate the liquid, gas and fluid phases free energy, but applied
perturbation theory for the crystal free energy. To mark the
difference with previous calculations, where phase equilibrium lines
had been calculated by perturbation theory both for the crystal and
for the fluid phase, we name our approach a hybrid method.

Phase boundaries between two phases (gas-liquid, fluid-solid etc.) are
obtained by imposing the standard conditions,
\begin{eqnarray} 
       \mu^{(1)}&=&\mu^{(2)} \label{equcon1}\\ 
       P^{(1)}&=&P^{(2)} \label{equcon2} 
\end{eqnarray} 
where $\mu$ is the chemical potential, i.e. the Gibbs free energy per
particle, $\mu=G/N$.

In those cases where SCOZA and perturbation theory are quantitatively
validated the equilibrium phase diagram is highly
accurate. As noted above, and shown in Figures \ref{fig1.8phdiag} and
\ref{fig6.0phdiag}, the SCOZA is well validated up to values of $b=9$.
For values of $b$ less than or equal to $6$, results are
indistinguishable in terms of phase equilibria from the best
simulations that have been carried out \cite{caccamo99,hagen94,shukla00}. 
Similarly, as we shall discuss later, the
perturbation theory rarely produces an error of more than $0.5\%$ in
the free energy of the solid phase, although this analysis is based on certain 
assumptions
about the perturbation series.  Combining these observations, we
believe that our phase diagrams are quantitatively accurate up to at
least $b=9$. Beyond that, we make no particular claim, except that we
expect that this hybrid method should still remain superior to the
typical theoretical approximations that have been applied
previously. Simulations have not been carried out beyond $b=9$.
 
\section{Glass Transition and Mode Coupling Theory} 
\label{sec:MCT} 
 
\subsection{Theory}

The study of the glass transition in colloidal systems has been one of
the most striking cases of verification of the current theories of
super-cooled liquids. Early experimental studies
\cite{pusey91,megen94} involved colloidal particles that are very
closely represented by hard spheres where only excluded volume effects
are important at high concentrations. Moreover, in contrast to simple
atomic liquids, it is possible to avoid the crystalline phase beyond
volume fractions of $49\%$ for sufficiently long periods of time to
study the glassy-type dynamical processes, and ultimately the
colloidal glass. The agreement between certain aspects of MCT and
experiments on colloids is quite satisfactory \cite{goetze99} and the
details of the time correlation functions are quite well
reproduced. It is widely believed that, in deeply supercooled
molecular liquids, the slow dynamics involves more complex dynamical
processes than those described by MCT, and there the theory becomes of
more qualitative applicability. Thus, the case of colloidal particles
is of some practical interest in applying this type of theory.
 
In fact, even for colloids, small
discrepancies appear in the comparison between experiments and
MCT. The most important is the value of the critical 
volume fraction for the hard-sphere arrest transition, the
experimental value being about $58\%$, while MCT predicts about
$52\%$. This is of little importance where the dynamical laws at the
hard-sphere transition are being compared between experiment and
theory.  Previous researchers have applied a shift to the transition
volume fraction, and then fitted the laws in this region
\cite{megen94}. Since the only current information on arrest driven
by attractive interactions is that provided by MCT \cite{A4}, there is
as yet no accepted manner in which we can correct the MCT curves. This
is somewhat inconvenient in the current context since for some parts
of the parameter space the equilibrium phase-diagrams are
quantitatively accurate, and it would be very satisfying to be able to
superimpose, without correction, the relevant MCT arrest curves.

 
We now briefly review the nature of MCT, and discuss the type of
information it yields. The MCT of super-cooled liquids describes the
non-ergodicity transition by a nonlinear integro-differential system
of equations for the normalized time correlation functions of density
fluctuations $\Phi(q,t)$. Apart from parameters entering from the
microscopic motion, the only input to the MCT equations is the
equilibrium wave-vector dependent structure factor of the system,
$S_q$. The glass transition lines can be identified by studying the
long time limit of the MCT equations, which determine the
non-ergodicity parameter of the system $f_q=\lim_{t\rightarrow
\infty}\Phi(q,t)$. An ergodic state is characterised by $f_q=0$. This
value is always a solution of the MCT long-time limit equations
\cite{goetze91}.  Thus, the glass transition appears as an ergodic to
non-ergodic transition for the system, where $f_q \neq 0$ solutions
arise. These points, thus, correspond to bifurcation singularities of
the MCT equations, and, depending on the number of control parameters
of the model, these can be of increasingly higher order, producing
interesting features of the arrested states diagrams.

A good quality $S_q$ is an important input for a good description of
the MCT arrest transition, as for the equilibrium phase diagram. In
the earliest discussions of colloidal systems with short-ranged
attractive interactions the Baxter interaction \cite{baxter,stell}, a
limiting case with an infinitely deep and zero-ranged SW attractive
potential, was discussed by a number of authors
\cite{fabbian99,chen,piazzapapers}.  Subsequent studies indicated that
the MCT equations are pathological for this interaction
\cite{foffi01}. The calculations were therefore extended to a
SW potential both in the PYA and the MSA \cite{A4}. Another solution
of the MCT equations was obtained using the Yukawa potential and the
MSA \cite{bergenholtz99}. Some common aspects emerged in these
works. However, the SW model at first appeared to give a richer
behaviour for the arrest transition curve. Thus, the system was shown
to possess a glass transition curve in the parameter plane $(\phi, k_B
T)$ the shape of this curve depending on the value of the SW parameter
$\Delta$ \cite{A4}. For narrow well-widths,     two branches of the glass
curve have been identified. These have been interpreted respectively
as a transition between a fluid phase and repulsion dominated glass
(this is the typical repulsive glass) and a transition between a
liquid and an attractive-interaction dominated glass (named the
attractive glass). The two branches join and for approximately $\Delta
\leq 4.11\%$, a glass-glass coexistence between the two different
types of glass appears. This coexistence line terminates in an
end-point, beyond which the non-ergodicity parameters become the same
for the two types of structures. The relevant singularity points, such
as the end-point or the point where the glass-glass transition line
reduces to a single point, are identified with higher order
singularities of MCT equations and lead to unusual logarithmic
dynamical relaxation laws \cite{A4,goetze91}.
 The mechanical properties of the system
have been also studied \cite{zaccarelli01} and they reinforce this
interpretation.  Earlier studies of the Yukawa potential did not
locate this glass-glass phenomenon \cite{bergenholtz99}, but it was
subsequently realized that the screening parameters
which had been studied were not large enough\cite{privatefuchs}, and further calculations seem to
give clear indications that both the SW and Yukawa potentials give the same
typical behaviour \cite{meridapaper}, implying that this does not
crucially depend either on the potential shape or the approximation
used for calculating the structure factor. Thus, it is now believed
that this glass-glass scenario, and the attendant dynamical laws, is
essentially a universal feature of the very short-ranged attractive 
potential.

We note in passing that the formation of two solid glass phases for
very short-ranged potentials should not be too surprising. In fact we
have earlier alluded to the fact that there are two crystalline phases
in the phase diagram of such potentials. We may typically view glasses
or arrested states as long-lived meta-stable states of the system that
have not been able to equilibrate to the nearby crystal, and that are
trapped in a restricted portion of phase space. In this sense we may
expect each crystal to have associated to it a particular glass-type.
Since one of the crystals in our phase diagram is ``attractive energy
dominated'', and the other ``repulsive energy dominated'', it is hardly
surprising that there should be two types of glass, dominated by the
two regimes of interaction.
 
A further comment on the relevance of these glass curves is that we
may see them as more than simply `transition curves'. If one reflects
more deeply on the nature of the equilibrium phase diagram, and the
thermodynamic states present in them, we recognize that at a deeper
level they are reflecting the fact that for those particular
parameters, the phase-space is dominated by a particular structure: a
crystalline, liquid, or gas structure.  The arrest curves carry
analogous information. Thus, in the vicinity of the arrest curve, we
may understand that most of phase space is becoming increasingly
inaccessible, and breaking into smaller regions that are
disconnected. That this may occur even for states that appear to have
static structures typical of liquids is the distinguishing feature of
glasses. Associated to this observation are those of dynamical slowing
and tendency to arrest during any phase-separation through which such
a glass curve passes. We shall argue in the conclusions that such
phenomena are relevant in protein crystallization.

\subsection{Experimental Studies}

We now discuss some of the particular experimentally determined
features that are associated with dynamical arrest driven by
attractive interactions. For example, Verduin and Dhont
\cite{verduin95}  determined a curve of structurally
arrested states in the phase-diagram of a system with short-ranged
attractive depletion interactions.  This locus, in some cases,
intersects the binodal line and is referred to as transient gelation
when observed in the spinodal region.  Interestingly, it was these
experimental\cite{verduin95} authors who were the first to comment that MCT might also
be applicable to cases where attractive interactions are important. As
we noted above, subsequent results of such calculations have been most
interesting
\cite{A4,zaccarelli01,bergenholtz99}. In addition a number of other
experimental programs involving particles with depletion interactions
have been published which offer many interesting insights for example
in \cite{grant93,meller99}. In particular, Poon {\it et al}
\cite{poon} have studied the arrest transition for systems where the
range of the interactions is short, and their more recent work in this
topic involves detailed connection to the theory described above
\cite{poonprivate}. Other systems may have certain advantages over the
depletion interaction system, but it is as yet too early to decide
this issue
\cite{bartsch,palberg,schaertlmainz,schurtenberger,bristol,bristol2,Levitzmanin}.
The results are typically quite promising, with some of these other
systems also exhibiting some of the phenomena predicted by the
theory. We may note in particular a more recent set of experiments
that are intriguing in that they make detailed predictions for the
correlation functions in a particular (re-entrant) part of the phase
diagram. Thus, concentration time correlation functions have been
observed in a polymer micelle system with a decay process much longer than the
usual stretched exponential, and the results have been well-fitted to a
logarithmic time-relaxation, \cite{mallamace00}, as predicted by the
theory\cite{gotzesjogren}.
 
The development of experimental understanding, and deepening of 
the theory of systems with short ranged potentials is really just 
beginning, and many experimental programs have now been commenced 
or reoriented in efforts to make progress. However early 
information indicates that the MCT type theory may be able to 
describe main elements of the principal phenomena, at least for 
reasonably high volume fractions, where it is possible to separate 
aging from dynamical arrest in a reasonably clean manner. 
 
\section{Are spinodals in Colloidal Systems meaningful?} 
\label{sec:spindal} 
Here we will take the liberty of raising a few issues in relation to
spinodal curves that are calculated via SCOZA, or indeed many other
typical liquid state theories. The reason that we make these comments
is that such curves should have a particular status for these
colloidal systems that are not relevant generally for molecular
systems. We note first that the spinodal curve is determined from the
condition that the curvature of the free-energy with respect to the
relevant density variable becomes zero, and that this corresponds to
the fluid phase becoming unstable as we lower the temperature. Between 
the binodal and the spinodal, the liquid and gas states in co-existence  may be
the global free-energy minima, but the fluid state remains
meta-stable.  Inside the region bounded by the spinodal, only the liquid 
and gas states in coexistence are
stable. Now it is well known
\cite{stanleybook} that the free energy is a convex function, and it
possesses only one minimum, and for some years now it has been
understood that the spinodal curve determined from approximate
theories (eg. mean-field theories) that consider two separate branches
of the free energy and then connect them, has no real scientific
basis.  Careful Monte Carlo simulations carried out in systems of
increasing size\cite{mcsimulations} have lead to the conclusion that
the spinodal curve shifts with the system size, merging with the
binodal in the limit of infinite systems. There is some loosely
defined kinetic phenomenon however\cite{mcsimulations,cahn-hilliard},
though even there it is not possible to define a spinodal curve, but a
cross-over regime where the kinetic mechanisms begin to change from
nucleation and growth to more collective phenomena. Interestingly
enough, when the particle size becomes large these more sophisticated
expectations are less relevant. Thus, it transpires that the relevant
parameter in this story is the ratio of the particle diameter to the
correlation length of the fluid. For very large particles such as high
molecular weight polymers, colloidal and other particles, the
microscopic length is so large that one has to be extremely close to the
critical point to see fully developed fluctuations beyond the
mean-field type ideas. Another consequence of this is that critical
exponents in such systems as proteins\cite{proteins} and
micelles\cite{micelles} have often been measured with mean-field values
because experiments were not performed in the true critical
regime. Similarly, the normal scepticism about the existence of a
spinodal curve should be less relevant here, and we may expect the
colloidal systems to exhibit quite reasonable spinodal behaviour. We
have therefore included the spinodal curve in our phase diagrams.\\
In concluding this section on dynamically defined objects in the 
phase diagrams, it is commented that we have chosen to plot the 
MCT curves through the meta-stable regions between binodals and 
spinodals. Again, in colloidal systems, for the reasons given 
above, it is to be expected that such curves would have meaning, 
whereas they would not be meaningful in molecular systems. 

\section{Phase-Equilibria and dynamical arrest lines} 
\label{sec:results} 
 
In this section we combine the results from the different 
techniques described above to exhibit the state of the system 
for given well depth. 

\subsection{Yukawa potential} 
\begin{figure} 
\centerline{\psfig{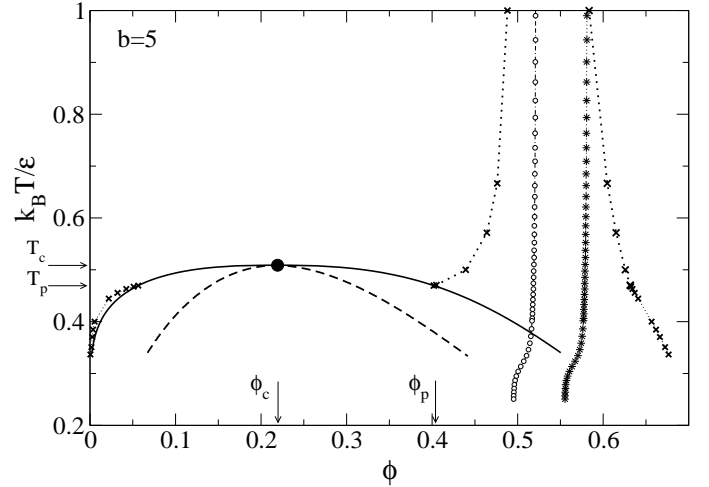}}
\caption{Phase diagram for the Yukawa fluid with screening parameter $b=5.0$. 
The crosses represent fluid-solid phase transition, the continuous
line is the binodal and the dashed one is the spinodal. The filled
circle is the critical point. The glass transition line as evaluated
for Mode Coupling Theory is also displayed (open circles). The glass
line shifted to obtain asymptotic value for $T\rightarrow\infty$ to be
the experimental packing fraction $\phi=0.58$ is presented
(stars). The subscripts $c$ and $p$ refer to the critical point and
the triple point respectively}
\label{fig5.0} 
\end{figure}

\begin{figure} 
\centerline{\psfig{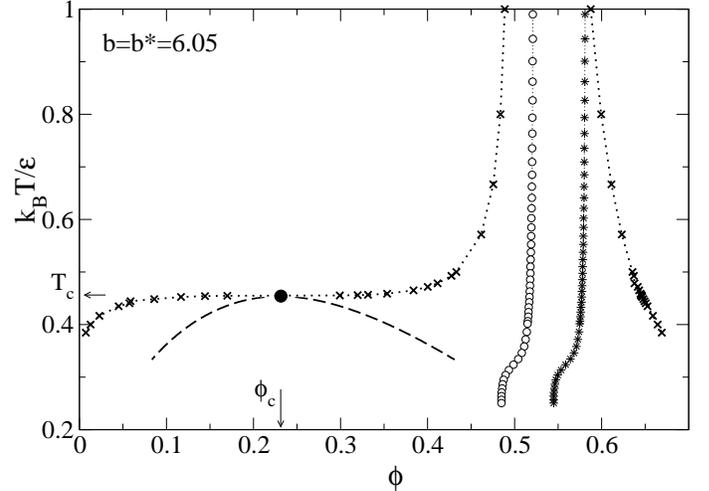}}
\caption{As in Fig. \ref{fig5.0} with $b=6.05$. The fluid branch of the fluid-crystal coexistence line now 
passes trough the liquid-gas critical point. The shifted glass line is not represented in this case.} 
\label{fig6.05} 
\end{figure}

In this section results for the Yukawa potential with the temperature
in units of the well depth
($k_B T/\epsilon$)
plotted against the volume fraction of the system, $\phi$ are
presented. The values of the well width are determined via the
screening parameter of the Yukawa $b$. The hard core radius is
fixed at unity, so all quoted lengths are in units of the hard core
diameter. We show results for values of $b=5$ (Fig. \ref{fig5.0}),
$b=6.05$ (Fig. \ref{fig6.05}), $b=7.5$ (Fig. 5), $b =30$
(Fig. \ref{fig30}), $b = 60$ (Fig. \ref{fig60}), and $b= 100$
(Fig. \ref{fig100}). For comparison we also show (Fig. \ref{SW0.03}) a
calculation for the square well system with $\Delta=0.03$.
 
We begin with the largest well width, corresponding to $b=5$, see
Fig. \ref{fig5.0}. Here the well width, considered say as the distance
of half amplitude of a Yukawa, is comparable to the particle hard-core
size. This is the typical situation that we are familiar with in
elementary phase diagrams of atoms and
molecules where van der Waals interactions predominate.  Thus,
we see the expected pattern of phase behaviour. Below the critical
temperature, the gas-liquid phase equilibrium occupies the greater
part of the low- and middle-range of densities, above the triple point.
The crystal is favoured at higher density, and the liquid- and
fluid-crystal boundary is nearly vertical, that is at fixed density,
reflecting the substantial absence of any energy scale in the
problem. The crystal is so tightly packed, and the attractions are so
spread out across the system that it is only the repulsive part of the
potential that is fundamental for crystallization. Indeed this is one of
the important ideas in traditional liquid state theory; that
attractions are not relevant to crystallization.
 
The asymptotic limits of solidification and melting boundaries of 
this coexistence at high temperatures reflect the hard-sphere 
limits respectively of $\phi\simeq 0.49$ and $\phi\simeq 0.55$, as 
expected from many simulations and theoretical observations 
\cite{gast83&86,alder}. The triple point temperature is labelled $T_p$. Within 
the crystal phase there is also the boundary for the dynamically
arrested state (the MCT transition line), again with high-temperature
asymptote of $\phi\simeq 0.52$(circles), the hard-sphere volume
fraction for the glass transition predicted by MCT. We recall that MCT
underestimates the glass transition packing fraction by about
$0.06$. To call attention on this shift we report the true MCT curve
(circles) as well as the MCT curve shifted by $0.06$ in packing
fraction using $*$ as symbol.  This boundary is also almost vertical,
again reflecting the fact that, for wider wells, the arrest transition
is driven essentially, at high enough temperature, by the repulsive
part of the potential. Thus, no attractive glass is observed for this
range of the potential.
 It has been shown theoretically \cite{gast83&86,evans},
experimentally \cite{exp} and by simulation \cite{hagen94}, that on
decreasing the range of the attractive potential the fluid-fluid coexistence
curve becomes meta-stable with respect to the fluid-solid one. This
means that for short enough potential ranges the liquid-gas critical
point is hidden in a phase separating region. This has two main
consequences.  The coexistence curve between the fluid and the solid
is very broad implying, for low temperatures, coexistence between a
very low density fluid and a high density solid (FCC).  Also, the
critical point becomes meta-stable with respect to gas-solid
co-existence. We will return to some of the possible implications of
all this later in the discussion. We have calculated  the
particular value of the screening parameter, $b$, at which the critical
point becomes meta-stable with respect to the low density fluid-gas
equilibrium, to be $ b^*=6.05$, and then presented the phase diagram at
that value (see Fig. \ref{fig6.05}). 
We note that slightly different values have been previously reported 
in the literature. Thus, Hagen and Frenkel quote the values $b^*=7.4$ 
using a Monte Carlo perturbation theory, and the value $b^*=6$ based on 
Gibbs Ensemble Monte Carlo (GEMC) \cite{hagen94}. Their GEMC value is 
very close to the value we have found. However a number of other 
values have also been quoted in the literature. Menderos and Navascues 
\cite{menderos} used a density functional approach to 
determine $b^*=8.25$, whilst Shukla \cite{shukla00} quotes a much
higher value of $b^*=13$. This latter value in particular is much
higher than previous ones and our value, and considering
that it is based on more extensive simulations than previous research
this might be a matter of concern. The issues in relation to the
accuracy of SCOZA have been addressed for modest values of $b$ in
Figures 1 and 2, where the liquid-gas phase diagrams were shown for
$b=1.8$ and $b=6$.  However, if we examine Table 4 and Figure 4 of
\cite{shukla00}, we can readily compare the simulations to the precise
predictions from SCOZA for selected values of the vapour and liquid
densities for a range of temperatures. For $b^*=6$, we find remarkable
agreement between SCOZA and the Shukla gas-liquid equilibria. Also,
whilst they are not quoted, we may estimate the critical temperature
for differing screening parameters from these same simulations, and
interpolation of others, and again conclude that the discrepancy
between what we find for $b^*$, and the Shukla result cannot arise
because of a difference between the SCOZA and GEMC results for
gas-liquid systems. The problem lies in the estimation of the crystal
free-energy, or entropy, or estimation of the triple point, the other
aspect of what we need to know to determine $b^*$. In
\cite{shukla00} the freezing transition location is determined by the
so-called one-phase entropic condition \cite{Giaquinta}. This
condition implies that the freezing density is essentially constant,
and almost unchanged for the range of $b$ values 1.8 to
10\cite{shukla00}.  In fact, in Figure \ref{fig6.05}, the phase
diagram for $b^*=6.05$, we do see rather significant deviation of the
freezing density in the vicinity of the critical temperature from its
high temperature limit. Indeed, comparing the other phase diagrams,
Figures 5-\ref{fig100}, we conclude that this variation is
intrinsic to the whole short-ranged scenario, since it is the prelude
to the splitting of the solid phase into two crystalline phases. The
true underlying discrepancy arises because the entropy criterion implies
that attractive forces are not important in the regime of
crystallization currently under discussion, whereas the 
perturbation theory implies that they are highly significant.  Both
approaches are approximations, but it is possible to estimate the
errors arising from the perturbation theory by considering the higher
order terms. 

Thus, for $b^* =6.05$, (here $k_BT_c^*/\epsilon= 0.454$
and $\phi_c^*= 0.230$) we can propose to bound the errors in the
perturbation theory of the crystal by reporting the ratios of the
second- to first-order terms in the perturbation series. In the
regime where the low-density fluid and crystal are in
equilibrium we find that the ratio of the second- to first-order terms
is approximately $0.0044$, whilst the first order-term has an absolute
value of $10.4$ \cite{comment}.  If this ratio represents a true
estimate of the errors, then the perturbation theory would appear to
be quite satisfactory, the curvature of the freezing curve genuine,
and the attendant flattening of the fluid side of the coexistence also
quite accurate.

In general by estimating the impact of such errors in perturbation
theory, we can estimate the shift of the fluid side of coexistence,
and thereby estimate errors. In so doing we find the typical error in
$b^*$ will be less than $1\%$ from this source. If we accept this
means of characterizing the error in perturbation theory, a matter
clearly based on the series being well-behaved at successive orders,
then we might conclude that the value of $b^*=6.05$, in agreement with
Frenkel's original calculation, is a good approximation. If this were
confirmed, then the calculations reported here for the overall
phase diagram are probably amongst the most accurate for modest values
of the range parameter, despite the fact that simulation is never
used. This is not the primary motivation of our paper, but it would be
an interesting way of approaching phase diagrams in future.
Despite these optimistic estimates, further more careful evaluations
are required by different methods to find a truly accurate
value. 

We should not imply that the value of $b^*$ is of such
crucial importance in the overall picture offered here. 
However, it does provide a useful check between
different researchers and methods of approximating the phase behaviour
in that its accurate estimation requires some satisfactory and
simultaneous treatment of gas, liquid and solid phases.
 
Finally, we note that 
ten Wolde and Frenkel have made some interesting comments in relation to the 
kinetic processes that might be expected in this regime
\cite{frenkelscience}.

In Fig.5 the case $b=7.5$ has been plotted. As noted by a
number of authors, the meta-stable gas-liquid phase equilibrium curve
has now flattened considerably, and the low-density fluid-crystal
coexistence (on the fluid side) has nearly the same slope. Increasing
curvature of the crystal side of the fluid-crystal equilibrium curve
is observed, arising from the increased influence of attractions on
the crystal.  The binodal line and the spinodal line in this case lie
completely within the region of fluid-solid phase separation, the
triple point has disappeared and the critical point is buried below
the flat part of coexistence curve between the low-density fluid and
the solid.  As we mentioned earlier, this meta-stable behaviour would
normally not be observable due to fluctuations, but for colloidal
systems, globular proteins and other large particles, we may expect to
observe such phenomena. Thus, on quenching such a system we might
expect to see rather a rich pattern of behaviour, depending on the
density that we quench at, and the depth of the quench. In particular,
it is noted that we should be able to see a meta-stable gas and a
liquid, the latter arresting into a glass at sufficiently low
temperature, because the glass curve crosses the binodal, and spinodal
at a finite temperature. We have earlier alluded to the idea that
critical fluctuations can play an important role in the formation of
crystals, for example protein crystals \cite{frenkelscience}. The
present screening parameter regime would exemplify this type of
phenomenon since here we have a meta-stable liquid and gas that are
critical (for large particles this should have some observable
lifetime).  The equilibrium phase diagram exhibits a fluid-crystal
co-existence, so we have the possibility of crystal nucleation and
growth phenomena in the presence of this meta-stable critical fluid,
and it is this matter that ten Wolde and Frenkel have discussed
\cite{frenkelscience}. However, we also note the point that for this
value of the screening parameter, as yet, the glass curve has not
begun to severely interfere with the gas-solid equilibrium curve, and
this is an additional advantage in the formation of crystals. This is
in distinction to subsequent phase diagrams where the glass curve 
extends across much of the space.
 
It is in fact worth reflecting on the shifted, more realistic
placement of the glass transition curve (stars). Thus, we see that at
high temperature, as expected, from $55\%$ to $58\%$ volume fraction
values, we have a crystalline state that is not interrupted by a glass
transition. However, as the temperature is lowered, the increased
importance of the attraction leads to the glass curve crossing the low
density fluid-crystal coexistence region, and beneath this, the
crystal may never form without the glass being an alternative long
lived state. This comment is relevant also to the case of $b=6.05$,
but for $b=7.5$, the curvature of the solidification curve has
increased greatly, so this effect is emphasized. This dramatic interruption
of the crystallization scenario will become more and more significant
as the range of the potential narrows, and will be an important theme
in our discussion.

Now we  turn our investigation to the case of very narrow potential
ranges. In Fig. \ref{fig30} we present the case $b=30$.  The two-phase
coexistence of fluid and solid now occupies a much larger portion of
the parameter space. The gas-liquid critical point is hidden well
below the fluid-solid coexistence line and on the crystal-fluid phase
boundary the effects of critical fluctuations will be much less. It is
interesting to note that the crystal side of the coexistence curve
exhibits a strong deviation towards higher densities. Thus the
coexisting solid will be much more dense that in the previous cases,
since the strong short ranged attraction is pulling the particles
closer at low temperature. This phenomenon is the precursor of a
solid-solid phase coexistence that in this case is still meta-stable
and lies within the sublimation curve.
 
The glass transition curve is most interesting. Of course, for $T 
\rightarrow \infty$, the curve asymptotes to the hard core value. 
The attractive forces at low temperature now begin to strongly affect
the curve so that it now turns sharply to the left, passing very close
to the submerged critical point. Thus, we may tentatively assign
$b^{**}=30.0$ as that value of the screening parameter at which the
submerged critical point becomes submerged by the glass, as well as by
the crystal-gas curve. The inherent inaccuracies in the MCT estimate,
alluded to before, may mean that the $b^{**}$ value may not be very
accurate. However the phenomenon is interesting. It means that the
meta-stable critical fluid is now competing with a glass
transition. It will transpire that the nucleation rate of these
fluid-crystal equilibria are very low, possibly due to the high
interfacial tension and therefore it is feasible that one may be able
to approach the liquid gas equilibrium and its nearby glass transition
without significant interference of the crystal.  The type of slowing
that would arise from a combination of critical slowing down and
glassy slowing down has not been discussed in the literature
previously. It would be an interesting problem. However,
the phenomenon discussed here would significantly affect the
possibility to form high-quality crystals, perhaps rendering it
essentially impossible. Even though the glass may eventually decay in
favour of the crystal, this will never lead to high quality
crystals. Of course, an alternative view of this situation is that
under these conditions it may be possible to make interesting
materials that have critical fluctuations frozen into the glassy
phase. Protein scientists wish to make good quality crystal, materials
scientists often wish to make interesting materials. Our comments are
applicable to both situations.
 
It is interesting also to note that in this case the glass curve
passes close to the gas-liquid critical temperature,but then dips and
intersects the binodal line below the critical density, passing
through the spinodal region. Such a scenario has been found by Verduin
and Dhont
\cite{verduin95} in experiments on colloidal systems.
We should stress that for low densities the glass curve itself may not
be reliable, as was discussed in \cite{zaccarelli01}. The situation in
relation to this point is, as yet, not settled.\\
We now discuss the case of extremely narrow wells. It is worth noting
that the previous two values of the screening parameters represent the
typical range of values accessed by those studying depletion-induced
attraction between colloidal particles \cite{poon} or globular proteins
\cite{piazza00,tardieu99}. The next set of phase diagrams 
(Fig.\ref{fig60} and Fig.\ref{fig100}) represent the limit of these
types of interaction, and may correspond to cases such as grafted
coatings on latex particles where we can access much narrower ranges
of potentials.\\
In Fig. \ref{fig60} and Fig. \ref{fig100} we present the cases $b=60$
and $b=100$ respectively. For numerical reasons, it is difficult to
extend fluid-solid coexistence curves in the low-density region of the phase
diagram. This is evident from the truncation of the phase diagrams, and
has no fundamental significance. As already noted 
in this paper, SCOZA has not been tested with simulations for
such narrow ranges of the attractive potential. On the other hand, the main 
phenomena 
appear using other methods, so at least these are expected
to be reliable. We return to this point at the end of this section.\\
In both figures we observe similar features. The spinodal line is now
buried deep in the sublimation curve as is the critical point.  A most
striking feature of these phase diagrams is the presence of a
solid-solid coexistence. This first-order phase transition was already
present for less narrow ranges but there it was meta-stable (see for
example the shoulder in the phase diagram for $b=30$). This phase
boundary represents the coexistence of two crystals with the same
lattice structure but different lattice spacing and consequently
different density\cite{iso-frenkel}.  It is terminated by a critical
point of the solid-solid coexistence. The origin of this co-existence
is interesting. The presence of short-ranged attractive interactions
causes competition with the hard-core repulsive interaction. The fact
that both have variations that occur on very short length scales means
that the system may be forced to `choose' between the 
attractive-dominated crystal and the repulsive-dominated crystal.\\
 In the low-density crystal region, i.e. $\phi < 0.65$, the
crystallization is dominated by entropic effect. In other words the
 the system chooses to optimize the entropy
to form an FCC structure. Increasing the density the particles become
closer and at some density they are forced to remain in the attractive shell
of their nearest neighbours. When this happens, there is a  decrease
in energy which leads to an
``attractive'' crystal. It is the energy that 
stabilizes the phase. It is indeed clear that such a phenomenon can be
present only if the range of the attractive potential is short
enough.  
We may note that an isostructural phase transition has been already
discussed theoretically for other kinds of potential characterized by
a short-range potential \cite{tejero} and indeed was also detected by
simulation \cite{iso-frenkel,evans}. It is almost certainly a genuine
phenomenon.  Here the isostructural phase transition is present for
both $b=60$ and $b=100$. Decreasing the range moves the critical
point of the transition to higher density, and indeed this is also in
agreement with the behaviour in simulations \cite{iso-frenkel}.  
It is interesting to note that it is possible to find a triple point
$T_p$ at which the two solids and the fluid coexist at the same
temperature.\\
In Fig. \ref{fig60} and Fig. \ref{fig100} the glass lines have been also 
plotted. They both tend to the hard-sphere limit for high 
temperatures, and bend towards low densities with decreasing 
temperature, as we have seen in the previous cases. For short enough 
interaction ranges the glass transition line does not pass close to 
the critical point. We note that for $b=100$ the glass curve appears to 
break into two branches, with an apparent discontinuity at that point 
marked $P_D$ in the figure. The low-density branch is called attractive 
glass, whilst the right-hand branch is called repulsive 
glass. For very short ranges, in other types of attractive potentials, 
we have located a glass-glass transition, a transition between two 
different type of glasses originated either by repulsion or by 
attraction \cite{A4}.  For the Yukawa fluid such a phenomenon also 
appears to be present, although it has not yet been investigated in 
detail \cite{privatefuchs,meridapaper}. \\
Both the results for the Yukawa potential considered here and those 
for the square-well fluid that we present next for comparison clearly show
that the distinction between the attractive solid and
repulsive solid becomes sharper as the range of the
potential becomes narrower. \\
To present in a coherent way the role
of the attractions on the crystal, glass and liquid-gas coexistence
line, we show in Fig. \ref{fig:Ttrends} the dependence of $T_c$ on $b$ and
both the glass transition temperature and the solid-fluid first-order
transition temperature at the critical packing fraction $\phi_c$. For
completeness, the inset shows $\phi_c(b)$.
\subsection{Square-well potential} 
Finally we discuss a single example for the SW model. By so doing we
wish to make the point that the main phenomena that have been
discussed above are independent of the details of the shape of the
potential, and are essentially universal.  We note, however, that the
SW phase diagram is not expected to be so quantitatively accurate as that 
of the Yukawa potential for
reason discussed in section {\ref{sec:solid}}.\\
The SW model was solved as
discussed in section \ref{sec:sw}. We discuss a case where the range
of the potential is very narrow, i.e. $\Delta=0.03$, and the result is
presented in Fig.\ref{SW0.03}. It is clear that the situation is very
similar to that for the Yukawa fluid, i.e. a solid-fluid phase
coexistence extends from high temperatures (where again it reaches the
correct hard sphere limit) expanding dramatically towards low and
high densities for low enough temperatures. An isostructural
solid-solid phase transition, with a critical point, is also present
in this case. Indeed a limited correspondence between different
potentials based on their general characteristics (effective core,
range and energy scale) has been recently proposed \cite{noro}, and it
is possible that this idea may have more general applicability.
  We hope to return to this more general concept of corresponding states at a later point \cite{us-unp}.\\ 
The glass line also has a very similar shape to the Yukawa fluid.  As
noted above, and in earlier publications, we have a glass-glass
transition, terminating in a higher order glass singularity,the A3
transition \cite{A4}. We believe that the presence of these two types
of glass, the attractive and repulsive glass, is the disordered
analogue of the presence of the two types of crystal discussed in some
detail above. In this case we may draw also an analogy between the
presence of the isostructural critical point and the presence of this
MCT singularity $A_3$ end-point at which the two glasses become
identical. Also as the well width gets larger, the crystal-crystal
critical point  vanishes. We believe that the
glass-glass analogous of this is the $A_4$ point \cite{A4,goetze91}.\\
It is quite reasonable to suppose that for every crystal, there should
be an analogous glass, and for every critical point of such a
crystal-crystal equilibrium there should be such an MCT singularity. It
will be interesting to explore this idea in future.  It has the appeal
of a potential general joint classification of equilibrium and glass
transitions.  %
 
\section{Conclusions} 
\label{sec:conclusions} 
We shall use the conclusions section of this paper for two purposes.  Thus
we shall attempt to sum up the practical conclusions of our
calculations, but, at the same time, try to considerably broaden the
discussion to make contact with the main experimental situations where
they might be useful. So far, we have focused the discussion quite
strongly on the narrowly defined consequences of studying a
short-ranged hard core attractive Yukawa potential, so we will begin
by summing up that aspect of the discussion.\\
Firstly, from the technical point of view, we have achieved a certain
success combining a good liquid-state method of calculation with
perturbation theory. The resulting hybrid method takes free-energies
from SCOZA for the gas and liquid states, and from perturbation theory 
for the crystal. 
We have indicated, in broad terms, how this overall
strategy could be applied to phase diagrams in general, and how it may
be qualified by checking of errors and relating these errors to shifts
in the phase boundaries. Given the potential to exploit powerful methods
of liquid-state theory, and the remarkable success of perturbation
theory for the ordered state when one chooses the correct zeroth-order
state, this may be a competitive manner in which to proceed for many
problems in future.\\
On the other hand, the methods to determine dynamical arrest (eg. MCT)
are not nearly so developed, despite their relative success in
colloidal science. In particular, the absolute values of the density 
and temperature at which the arrest
takes place is not correct.  This should not be
surprising. Equilibrium theory has had the benefit of many more years
of development, and much more effort devoted to bring it to this
level of achievement whereas for the dynamical arrest alternative
routes or different approaches have been developed only recently
\cite{europhy}. However, this aspect is quite inconvenient. \\
As we try to apply this theory to more realistic situations in
colloids, materials and biology, we see increasingly this important
motif of a competition between the equilibrium Boltzmann view of the
matter, and the dynamically arrested aspects.  So far, these two
fields have developed somewhat in independent manners. However from
the point of view of these practical topics in nature, there is no
distinction, and it is often not the separate behaviours, but the
interplay and competition between them that is primary to the
scientific issue. This paper is one of the first attempts to make
connections between these phenomena, but one can clearly see the
limitations.  For greater insight into these topics, we will have
to address the possibility to both improve systematically the methods
of studying arrest transitions, and also their consistency with
equilibrium transitions. This must represent one of the important
technical challenges in coming years.\\ From a broader perspective, we
have shown that when the range of the potential becomes short in
comparison to the core size, the subtle interplay between entropy and
energy begins to change its character. The range of densities over
which configurational entropy is relevant is much reduced, and one
begins to lose the liquid state, in favour of crystals, or arrested
glassy states. The reasons have been discussed at the beginning of the
paper. In essence they amount to the fact that to retain the benefits
of short-ranged attractions the particles must not depart too much
from their typical inter-particle distance, or they are no longer in
their mutual attractive well. This loss of freedom of motion, and
restriction of favourable configurations, leads to a lowered
configurational entropy. Another way of expressing the same idea is
that the short-ranged potential leads to the loss of easy fluctuations
that can open the cage of neighbouring particles that trap a central
particle. The probability of finding such an `opening' of the cage is
much reduced, and the time during which a particle is localized by its
neighbours increases, diverging at the arrest transition to form the
`attractive glass' that we have discussed. As the range of the
potential narrows, the means of egress permitted to the particles is
further limited, and the attractive glass becomes more favoured.  This
glass is therefore an effective competitor to the liquid and
crystalline phases of the system, and this is reflected in the fact
that the fluid phase is eventually erased by the glassy phase, and the
critical point submerged underneath the curve of arrest
transitions. This aspect should not be confused with the equilibrium
phase diagram, although it is interesting that there are many
parallels between the two. \\ As we discussed before, when the range
of the potential becomes very short, the competition between
entropy and energy is responsible for the formation of two distinct
crystal phases.  Coming from the fluid side, a crystalline phase
dominated by repulsion is present. This state arises from the fact
that the, at such densities, the entropy of a system made of particles
free to move only in their own Weigner-Seitz cell of a FCC structure
is larger then that of the metastable fluid . We named this phase a
repulsive crystal by analogy with the glass. On increasing the density,
this crystal becomes unstable and makes a transition to a smaller
Weigner-Seitz cell, so that particles mainly stay within
the attractive well. The structure of this phase, that we named
attractive crystal, is a more compact FCC crystal that optimizes the
free energy by means of reducing the potential energy.  The result at
this point is that the crystal structure may adopt two different
states. The situation is analogous to the case of a gas-liquid phase
separation where the gas (low density) phase optimizes the entropy and
the liquid (high density) has a much lower energy. \\ The implications
of all of this are profound for practical situations. For reasons
given earlier, many systems such as colloids, globular proteins,
``buckyballs'', nano-particles, pre-ceramic particulates, and others
have this property of a short-ranged attraction. In all these cases
``precipitation'', ``gellation'', ``glassification'', or solidification are
frequently the commonly observed outcome. In cases where we
consciously seek to make such a state this is satisfactory, and it
remains only to adjust the potential to have sufficiently short range
to obtain the required properties of the solid. However, for cases
such as globular proteins, and nano-scale or meso-scale ordered
materials with prescribed optical properties the situation is quite
different. Here we seek to make a crystal. In fact, reviewing the
phase diagrams in Figs. 5-\ref{fig100}, we can see why the crystal
is hard to access. If we work to the right side of the glass curve
(point marked $A$ in Fig. \ref{fig30}), universally we may expect to
fall into the glass state; there is essentially no choice. Since the
glass curve moves to low density, this is a serious
restriction. However, we may choose to work within the two-phase 
low-density fluid-crystal co-existence regime, but to the left of the
glass curve (point marked $B$ in Fig. \ref{fig30}). The outcome is then
a question of kinetic control, and will not be completely settled by
diagrams such as we are drawing. However we can make some educated
comments. Thus, if we work in the two-phase regime of gas-solid ($B$),
we may nucleate and grow crystals. Whether the proximity to a
meta-stable critical point is advantageous or not, as discussed by
Hagen and Frenkel \cite{hagen94} is not our primary concern here,
though this is an interesting proposition. The broader point is that
by nucleating to the left of the glass curve one may enter the
crystalline region (thus form a macroscopic crystal) by a route not
described by the ``adiabatic'' description here, and thereby avoid
some of the complications of the glass. This renders the formation of
crystal at least feasible, although where the glass curve runs through
the two-phase region, it will remain difficult to form truly large
high quality crystals. From this region where it is possible to
crystallize, one should also exclude the two phase region, whether it
is meta-stable or not, since the partial (micro-)phase separation,
crystallization and glassification, all competing dynamically is
unlikely to produce a good crystal also. This leaves only that region
bounded to the right by the glass curve, to the bottom by phase
separation, and to the top by the gas side of the gas-crystal
phase-coexistence as a likely candidate for forming good
crystals. This is interesting. It leads us to suppose that 
for a fixed short range of the potential
there is a
``practical crystallization region'' 
in the temperature-density plane, 
irrespective of the specific features of the equilibrium phase diagram.
However, more
importantly, there is also a limited regime of interaction ranges
where such a slot is significantly large enough to be accessed
experimentally. We believe that much of the discussion that has taken
place in the last few years in the literature in relation to protein
crystallization
\cite{piazza00,mushol95,wilson} is almost
certainly heading in the correct direction. Thus, Mushol and Rosenberg
\cite{mushol95} show in their Figure 12 the typical situation for a
globular protein phase diagram. They exhibit a phase diagram that has
(a) two metastable liquid phases in equilibrium, a more and less
dilute phase of protein (gas-liquid in our language), (b) a 'gellation
curve' that we associate with the glass curves in our work, and (c) a
fluid-crystal coexistence regime. They name the `good' regime for
crystallization zone I, and the others II and III. We have also
essentially partitioned our phase-diagram into the same types of
zones, and concluded that this gas-crystal region described above by
the glass, binodal and gas side of the coexistence boundary (zone I in
their language) would be the most favourable for formation of
crystals. For the sake of comparison in one of our figures
(Fig.\ref{fig30}) that has a range typical of globular proteins, we
have marked regions I-III in analogy with ref.\cite{mushol95}.\\ We
would argue that the present work, with its phase diagrams, and
discussion of control parameters is the quantitative expression of
these ideas that have arisen in the protein crystal literature. This
is potentially encouraging, since it opens the possibility to make
more quantitative study of these systems.\\
However, perhaps the most promising directions involve the study of
the current model, and underlying ideas in more depth to see what
independent kinetic routes exist to the formation of crystals. We have
seen how these phase diagrams are indicative of the kinetic behaviour,
but we believe that there will be much more significant insights as to
how to form high quality crystals in these regimes if a deeper
understanding of kinetics is acquired in future. In particular the
glass analogy seems promising as a means to characterize the more
confined phase space experienced by these systems. Aging, and kinetic
phenomena in general, is an arena that is growing in importance
\cite{kurchan,fdtviolation,latz,aging,sciortino} and may offer significant
advances. The traditional viewpoint of activated processes, and simple
kinetic processes is without doubt incomplete in the limit where we
approach the rather confined phase-spaces characterized by approach to
a glass transition. The realization that we are in a `glassy' scenario
may well assist in the development of new theories of kinetics of
crystallization more appropriate for such questions.\\
In any case, one can hardly doubt the high degree of practical 
significance that  kinetic phenomena associated with short-ranged 
potential systems will have in the coming few years. Given that we 
discuss a model potential that is only slightly different from 
those long considered in liquid-state 
theory, we must be intrigued by the  novelty in 
supposedly simple situations.

\acknowledgements 
{We acknowledge helpful comments by J.Bergenholtz and M.Fuchs who first
pointed out to us that the Yukawa model would also possess an $A_3$ and
$A_4$ scenario, but at much large screening parameter then their
previous studies. Also, the most helpful advice and comments of
R. Evans and D. Frenkel are gratefully acknowledged.\\ The work in Rome
is supported by PRIN-2000-MURST and PRA-HOP-INFM, and the work both in
Rome and in Dublin is supported by COST P1.\\
The work at Stony Brook was supported by the Division of Chemical Sciences,
Office of Basic Engineering Sciences, Office of Energy Research, U.S.
Department of Energy.}

\section*{Figures} 
 
\begin{figure} 
\centerline{\psfig{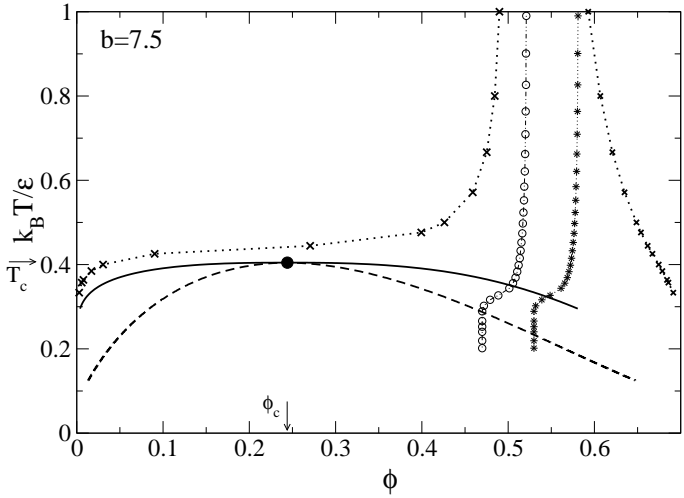}}
\label{fig7punto5}   
\caption{As in  Fig.\ref{fig5.0}  with $b=7.5$.} 
\end{figure}

\begin{figure} 
\centerline{\psfig{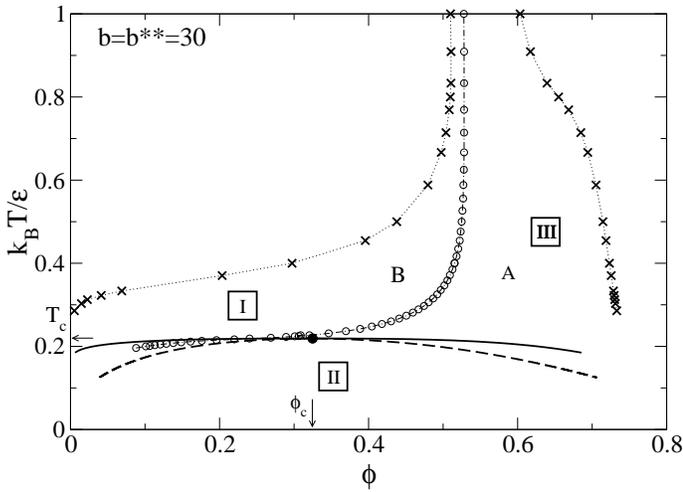}
           } 
\caption{As in Fig.\ref{fig5.0} with $b=30$. For the points $A$ and $B$ see the text. At this $b$
  value the glass line passes trough the metastable liquid-gas critical point. The labels I, II and III are chosen by 
analogy with the proposition of Mushol and Rosenberg\protect\cite{mushol95}. See text for details.} 
\label{fig30} 
\end{figure}

\begin{figure} 
\centerline{\psfig{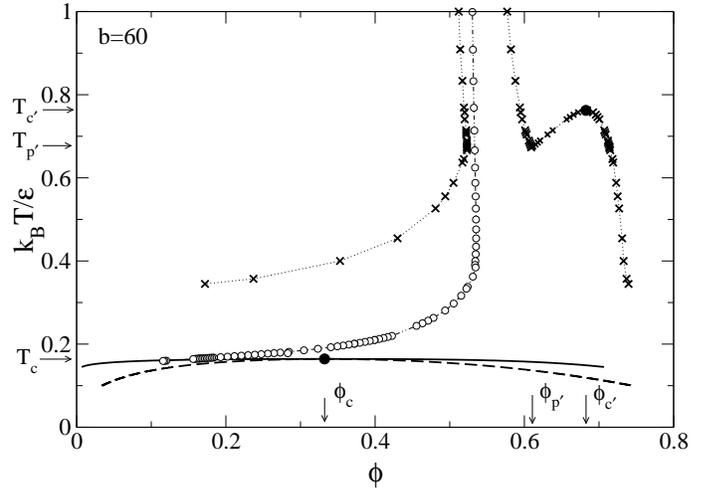}
           }  
\caption{As in Fig.\ref{fig5.0} with $b=60$. The subscripts $c'$ and $p'$ refer now to the the critical and triple points of the solid solid transition.The shifted glass line is not represented in this and in the following cases. } 
\label{fig60} 
\end{figure}

\begin{figure} 
\centerline{\psfig{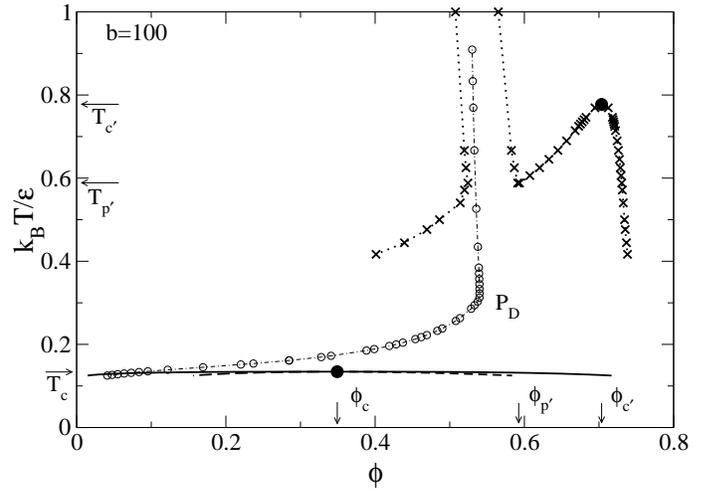} 
            }
\caption{As in Fig.\ref{fig5.0} figures for $b=100$.} 
\label{fig100} 
\end{figure} 
 
\begin{figure} 
\centerline{\psfig{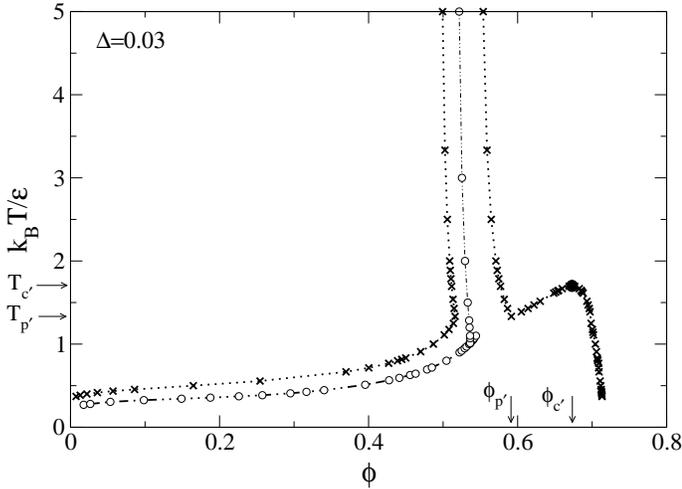}
           }  
\caption{Phase Diagram for the SW model for $\Delta=0.03$. The 
crosses represent the solid-fluid phase coexistence and the set of
open circle is the glass line. Note the solid-solid coexistence on the
high densities side of the phase diagram: its critical point is
labelled by a filled circle. The position of the liquid-solid-solid
triple point is also displayed $(\phi_{p'},T_{p'})$}
\label{SW0.03} 
\end{figure} 
 
 \begin{figure} 
\centerline{\psfig{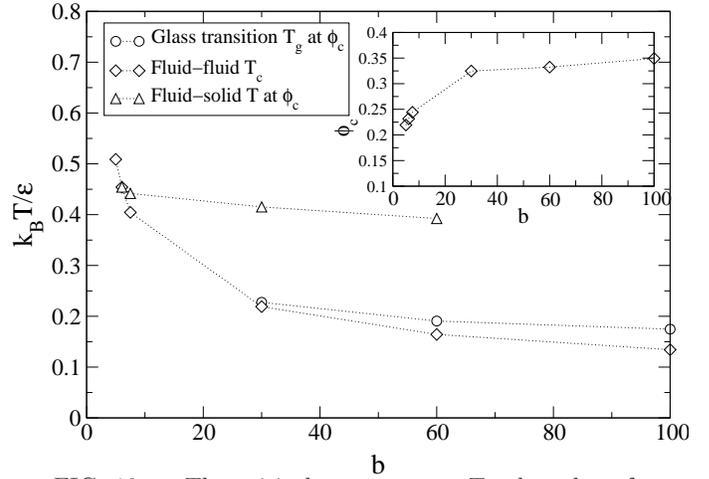}
           }  
\caption{
The critical temperature $T_c$ plotted as function of the screening
parameter $b$.  The glass transition and the solid-fluid coexistence
temperature at the critical packing fraction, $\phi_c$, are also
displayed. For completeness in the inset the $\phi_c$ as a function of
$b$ is also shown.}
\label{fig:Ttrends} 
\end{figure}

\end{multicols} 
 

\begin{references} 
 
\bibitem{piazza00} 
         R. Piazza 
         Current Opinion in colloid and interface Science 
         {\bf 5}, 38 (2000). 

\bibitem{mushol95} 
         M. Mushol and F. Rosenberg , 
         J. Chem. Phys. {\bf 103} 10424 (1995). 
 
\bibitem{lekkerkerker} 
         H. N. W. Lekkerkerker, W. C. Poon, P. N. Pusey, A. Stroobants 
         and P. B. Warren, Europhys. Lett. {\bf 20}, 559 (1992). 
 
\bibitem{verduin95} 
         H. Verduin and J. K. G. Dhont, J. Coll. and Interf. Sci. 172, 
         425 (1995). 
 
\bibitem{poon} 
    P. N. Pusey, A. D. Pirie and W. C. K. Poon, Phys. A {\bf 201}, 
    322 (1993); P. N. Pusey, P. N. Segr\`{e}, O. P. Behrend, 
    S. P. Meeker and W. C. K. Poon, Phys. A {\bf 235}, 1 (1996); 
    W. C. K. Poon, Curr. Op. Coll. Int. Sci. {\bf 3}, 593 (1998). 
 
\bibitem{bartsch} 
         E. Bartsch, V. Frenz, J. Baschnagel, W. Sch\"artl, H. Sillescu 
         J. Chem. Phys. 106, 3743 (1997); A. Kasper, E. Bartsch, 
         H. Sillescu, Langmuir 14, 5004 (1998). 


\bibitem{buckyballs}  M.H.J.  Hagen, E.J. Mejer, G. Mooji,D. Frenkel,H.N.W Lekkerkerker Nature  {\bf 365} 425 (1993). 


 
\bibitem{debenedetti01} 
         P. G. Debenedetti and F. H. Stillinger, 
         Nature {\bf 410} (2001) 259, 
         and references quoted therein. 
 
\bibitem{ISentropy} 
         F. Sciortino, W. Kob and P. Tartaglia, 
         {\em Phys. Rev. Lett.}  {\bf 83} (1999) 3214. 
 
\bibitem{A4} 
         K.A. Dawson, G. Foffi, M.Fuchs, W.G\"{o}tze, F. Sciortino, 
         M. Sperl, P. Tartaglia, Th. Voigtmann and E. Zaccarelli, 
         Phys. Rev E, {\bf 63}, 011401 (2001). 
 
\bibitem{zaccarelli01} 
          E.Zaccarelli, G.Foffi, P.Tartaglia, F.Sciortino and K.A.Dawson 
          Phys. Rev E, {\bf 63}, 031501 (2001). 
 
\bibitem{iso-frenkel} 
     P. Bolhuis and D. Frenkel, Phys. Rev. Lett. {\bf 72}, 2211 
     (1994); P. Bolhuis, M. Hagen and D. Frenkel Phys. Rev. E, 
     {\bf 50}, 4880 (1994). 
 
\bibitem{tejero} 
    C. F. Tejero, A. Daanoun, H. N. W. Lekkerkerker and M. Baus, 
    Phys. Rev. Lett. {\bf 73}, 752 (1994); Phys. Rev. E {\bf 
    51}, 558 (1995). 

\bibitem{evans} 
         M.Dijkstra, J.M. Brader and R.Evans 
         J.Phys.:Condens. Matter, {\bf 11}, 10079 (1999). 

\bibitem{gast83&86} 
         A.P.Gast, W.B. Russell and C.K. Hall, 
         J. Colloid Interface Sci. {\bf 96}, 1977 (1983); 
         A.P.Gast, W.B. Russell and C.K. Hall, 
         J. Colloid Interface Sci. {\bf 109}, 161 (1986). 
 
\bibitem{hagen94} 
         M. Hagen and D. Frenkel, 
         J. Chem.Phys, {\bf 101}, 4093 (1994). 

\bibitem{private-evans}  R. Evans private communication. 


\bibitem{goetze91} 
         W. G\"otze in 
         {\it Liquids, Freezing and Glass Transition} 
         edited by J.P. Hansen , D. Levesque D, and J. Zinn-Justin  
         (Amsterdan: North-Holland) p~287, 1991. 
 
\bibitem{bergenholtz99} 
         J. Bergenholtz and M. Fuchs, Phys. Rev. E {\bf 59}, 5706 
         (1999). 
 
\bibitem{fabbian99} 
         L. Fabbian, W. Goetze, F. Sciortino, P. Tartaglia and F. Thiery, 
         Phys. Rev. E {\bf 59}, R1347 (1999). 

\bibitem{pini3} 
         D. Pini, G. Stell and N. B. Wilding, 
         Mol. Phys. {\bf 95}, 483 (1998). 

\bibitem{caccamo99} 
         C. Caccamo, G. Pellicane, D. Costa, D. Pini, G. Stell 
         Phys. Rev. E {\bf 60}, 5533 (1999). 

\bibitem{sear99} 
         R. Sear, J. Chem. Phys. {\bf 111} 4800 (1999).
 
\bibitem{pini01} 
         D. Pini, G. Stell and N. B. Wilding 
         J. Chem.Phys {\bf 115}, 2702 (2001). 
 
\bibitem{carnahan69} 
         N. F. Carnahan and K. E. Sterling 
         J. Chem. Phys {\bf 51}, 635 (1969). 

\bibitem{waisman} E. Waisman, 
         Mol. Phys. {\bf 25}, 45 (1973).
 
\bibitem{hansen86} 
         J.P. Hansen  and I.R. McDonald  1986 
         {\it Theory of Simple Liquids} 
         (London: Academic Press). 
 
\bibitem{shukla00} 
        K. P. Shukla, J. Chem.Phys, {\bf 112}, 10538 (2000). 
 

\bibitem{rascon} 
        C. Rascon, L. Mederos and G. Navascues, Phys. Rev. Lett. {\bf
        77}, 2249 (1996).
 
\bibitem{PY} 
         J.K. Percus and G.J. Yevick 
         Phys. Rev., {\bf 110}, 1 (1958). 
 
\bibitem{baxter68} R. J. Baxter, 
         Aust. J. Phys. {\bf 21}, 563 (1968). 
 

 
\bibitem{barker} 
         J. A. Barker and D.J Henderson, 
         J. Chem.Phys {\bf 47}, 2856 (1967). 
 
\bibitem{numrecipes} 
         W. H. Press, S. A. Teukolsky, W. T. Vetterling, 
         B. P. Flannery {\it Numerical Recipes in C} 
         2nd edition (January 1993) Cambridge Univ Press). 
 
\bibitem{alder} 
         B. J. Alder, W.G. Hoover and D.A. Young, 
         J. Chem.Phys {\bf 49}, 3688 (1968). 

\bibitem{gasser01} 
	U.Gasser, E. Weeks, A. Schofield, P.N. Pussey and D.A. Weitz
	Science {\bf 292 } 258 (2001)

\bibitem{auer01} S. Auer and D. Frenkel {\bf 409 } 1020 (2001)

\bibitem{pronk01} 
         S. Pronk and D. Frenkel
         J. Chem.Phys {\bf 110}, 4589 (1999). 
 
\bibitem{hall} 
         K. R Hall 
         J. Chem.Phys {\bf 57}, 2252 (1971) 

\bibitem{frenke-pri} 
         D. Frenkel, 
         private communication. 
 
\bibitem{kinc} 
          J. M. Kincaid and J.J. Weis, 
          Mol. Phys. {\bf 34}, 931  (1977). 
 
\bibitem{VW} 
         L. Verlet and J.J. Weis
         Phys. Rev A, {\bf 5}, 939 (1972). 
 
\bibitem{pusey91} 
         P. N. Pusey in 
         {\it Liquids, Freezing and Glass Transition} 
         edited by Hansen J -P, Levesque D, and Zinn-Justin J 
         (Amsterdan: North-Holland) p~763, 1991. 
 
\bibitem{megen94} 
         W. van Megen and S. M. Underwood, 
         Phys. Rev. Lett. {\bf 70}, 2766 (1993), 
         Phys. Rev. E {\bf 49}, 4206 (1994). 
 
 
\bibitem{goetze99} 
         W. G\"otze 
         J. Phys.: Condens. Matter {\bf 11} A1, (1999). 
 
\bibitem{baxter} 
         R. J. Baxter, 
         J. Chem.Phys, {\bf 49}, 2770 (1968). 
 
\bibitem{stell} G. Stell, J. Stat. Phys. {\bf 63}, 1203 (1991).    
 

 
\bibitem{chen}   
        Y.C. Liu, S.H. Chen and J.S. Huang, Phys. Rev. E {\bf 54}, 
        1698 (1996). 
 
 
\bibitem{piazzapapers} R. Piazza, V. Peyre and V. Degiorgio  
                Phys. Rev. E {\bf 58} R2733 
         (1998) 
 
 
\bibitem{foffi01} 
         G. Foffi, E.Zaccarelli, F. Sciortino, P.Tartaglia and K.A. Dawson 
         J.Stat.Phys, {\bf 100}, 363 (2000). 

 
\bibitem{privatefuchs} J. Bergenholtz and M. Fuchs, private communication. 
 
\bibitem{meridapaper}  
        K. A. Dawson, G. Foffi, G. D. McCullagh, F. Sciortino, 
        P. Tartaglia and E. Zaccarelli, submitted to 
        J. Phys. Cond. Matt. (2001). 
 
\bibitem{grant93} 
         M. C. Grant and W. B. Russel, Phys. Rev. E {\bf 47} 2606 
         (1993). 
 
\bibitem{meller99} 
         A. Meller, T. Gisler, D. A. Weitz, and J. Stavans, 
         {\it Langmuir} {\bf 15}, 1918 (1999). 

\bibitem{poonprivate} W. Poon, private communication.

\bibitem{palberg} S. Neser, C. Bechinger,P. Leiderer,T. Palberg Phys. Rev. Let. {\bf 79}, 2348 (1997); 

\bibitem{schaertlmainz} W. Schaertl and C. Roos  Phys. Rev. E {\bf 60} 2020
         (1999). 
              

\bibitem{schurtenberger} S.H. Behrens, M. Borkovec, P. Schurtenberger 
	Langmuir {\bf 14}, 1951 (1998) 

\bibitem{bristol} S. Henderson, S. Mitchell and P. Bartlett 
		Colloid Surface A,  {\bf 190}, 81 (2001)
 

\bibitem{bristol2} J.R. Weeks , J.S.van Duijneveldt, B. Vincent 
 		J. Phys. Cond. Matt. {\bf 2}, 2227 (2000).

\bibitem{Levitzmanin} S. Meyer , P. Levitz and A. Delville, J. Phys. Chem B {\bf 105}, 9595 (2001);   I. Grillo, P. Levitz and T. Zemb, Eur. Phis. J E {\bf 5}, 377 (2001) 
 
\bibitem{mallamace00} 
         F. Mallamace, P. Gambadauro, N. Micali, P. Tartaglia, 
         C. Liao and S.H. Chen, 
         Phys. Rev. Lett. {\bf 84}, 5431 (2000). 

\bibitem{gotzesjogren} 
        W.~G\"otze and L.~Sj\"ogren, 
       J. Phys. Cond. Matt. {\bf 1}, 4203 (1989).

\bibitem{stanleybook} H.E. Stanley, 
                      {\it Introduction to Phase Transitions and Critical Phenomena} 
 		Oxford University Press (1997)

\bibitem{mcsimulations} 
J.D. Gunton and M. Droz,{\it Introduction to the dynamics of metastable and unstable phases} Lecture Notes in Physics, Vol. 183, (1983),
ed. J. Zittartz (Springer-Verlag, Berlin).


\bibitem{cahn-hilliard} 
J.D. Gunton, M. San Miguel and P.S. Sahni, in {\it Phase Transition and Critical Phenomena, VOl. 8} edited by D. Domb and J.L. Lebowitz (Academic, London 1983).

\bibitem{proteins}  
     P. Schurtenberger, R.A. Chamberlin, G.M. Thurston, J.A. Thomson and G.B. Benedek, Phys. Rev. Lett. {\bf 63}, 2064 (1989);
	B.M. Fine, J. Pande, A. Lomakin, O.O. Ogun and G.B. Benedek, 
	Phys. Rev. Lett.  {\bf 74}, 198 (1995) 

\bibitem{micelles} M.E. Fisher,  Phys. Rev. Lett.  {\bf 57}, 1911 (1986)

\bibitem{exp} 
         S.M.Illett, A.Orrock, W.C.K. Poon, P.N.Pusey 
         Phys. Rev E, {\bf 51}, 1344 (1995); 
         W.C.K. Poon, A.D. Pirie and P.N. Pusey 
         Faraday Discuss., {\bf 101}, 65 (1995); 
         W.C.K. Poon,{\it Phys. Rev E} {\bf 55}, 3762 (1995) 
 

\bibitem{menderos} L. Menderos and G. Navascues, J. Chem. Phys {\bf 106}, 9841 (1994)

\bibitem{Giaquinta} P.V. Giaquinta, Physica A {\bf 187}, 145 (1992).

\bibitem{comment}
	We estimated this ratio for different values of the screening
	parameter and for different temperatures. The largest value is
	found to be less than $0.025$ but the typical value is much
	smaller.  It decreases for high density. This agrees with
	Stell and Penrose comment that such an expansion should be
	exact at the first order in the close packing
	limit\protect\cite{stell84}.
 

\bibitem{frenkelscience} 
         P. R. ten Wolde and D. Frenkel 
         {\it Science} {\bf 277}, 1975 (1997). 
 
 
\bibitem{tardieu99} A. Tardieu, A. Le Verge, M. Malfois, F. Bonnete, S. Finet, M. Ries-Kautt  and L. Belloni 
				J. Cryst. Growth {\bf 196}, 193 (1999) 

\bibitem{noro} 
         M.G. Noro and D. Frenkel, 
         J.Chem.Phys. {\bf 113}, 2941 (200). 

\bibitem{us-unp}  G. Foffi, E.Zaccarelli, F. Sciortino, P. Tartaglia and
	K.A. Dawson {\it unpublished}.

\bibitem{europhy}       
	E.Zaccarelli, G. Foffi, F. Sciortino, P. Tartaglia and
	K.A. Dawson {\em Europhys. Lett.} {\bf55}, 139 (2001);
	E. Zaccarelli, G. Foffi, P. De Gregorio, F. Sciortino,
	P. Tartaglia and K. A. Dawson {J. Phys.: Condens. Matter, \ \
	submitted}

\bibitem{wilson}
	A. George and W.W. Wilson, Acta Crystallogr. D {\bf 50} 361 (1994). 


\bibitem{kurchan} 
        J-P. Bouchad, L. Cugliandolo, J. Kurchan and M.  Mezard,
        Physica A {\bf 226}, 243 (1996).

\bibitem{fdtviolation}
        W. Kob and J.-L. Barrat, Phys. Rev. Lett. {\bf 78}, 4581
        (1997); G. Parisi, Phys. Rev. Lett. {\bf 79}, 3660 (1997).

\bibitem{latz}
        A. Latz, J. Phys. Cond. Matt. {\bf 12}, 6353 (2000);
        cond-mat/0106086.


\bibitem{aging}
        P. De Gregorio, F. Sciortino, P. Tartaglia, E. Zaccarelli and
        K. A. Dawson, Physica A (in press), 2001.
     
\bibitem{sciortino}
        F. Sciortino and P. Tartaglia, J. Phys. Cond. Matt {\bf 13},
        9127 (2001).


\bibitem{stell84} 
         G. Stell and O. Penrose,  
         {\it Phys. Rev. Lett.} {\bf 52},  85 (1984).

\end{references}
\end{document}